\documentclass[11pt, oneside]{article}   	
\usepackage{jheppub}
\usepackage{amsmath}
\usepackage{amssymb}
\usepackage{amsfonts}
\usepackage{amsthm}
\usepackage{amsbsy}
\usepackage{array}
\usepackage{mathtools}
\usepackage[usenames,dvipsnames]{xcolor}
\usepackage{subcaption}
\usepackage{dsdshorthand}
\usepackage{graphicx}
\usepackage[vcentermath]{youngtab}
\usepackage{tikz}
\usepackage{float}
\usepackage{verbatim}
\usepackage{mathtools}
\usepackage{subcaption}
\usepackage{xcolor}
\usetikzlibrary{snakes}
\usetikzlibrary{decorations}
\usetikzlibrary{trees}
\usetikzlibrary{decorations.pathmorphing}
\usetikzlibrary{decorations.markings}
\usetikzlibrary{external}
\usetikzlibrary{intersections}
\usetikzlibrary{shapes,arrows}
\usetikzlibrary{arrows.meta}
\usetikzlibrary{calc}
\usetikzlibrary{shapes.misc}
\usetikzlibrary{decorations.text}
\usetikzlibrary{backgrounds}
\usetikzlibrary{fadings}
\usepackage{tikz}
\usetikzlibrary{patterns}
\usetikzlibrary{positioning}
\usetikzlibrary{tikzmark,calc,arrows,shapes,decorations.pathreplacing}
\tikzset{
        cross/.style={cross out, draw=black, minimum size=2*(#1-\pgflinewidth), inner sep=0pt, outer sep=0pt},
	branchCut/.style={postaction={decorate},
		snake=zigzag,
		decoration = {snake=zigzag,segment length = 2mm, amplitude = 2mm}	
    }}

\newcommand{\bea}{\setlength\arraycolsep{2pt} \begin{eqnarray}}
\newcommand{\eea}{\end{eqnarray}}
\def\ft#1#2{{\textstyle{\frac{\scriptstyle #1}{\scriptstyle #2} } }}
\def\fft#1#2{{\frac{#1}{#2}}}

\newcommand{\baa}{\begin{align}}
\newcommand{\eaa}{\end{align}}
\newcommand{\eq}[1]{\begin{equation}#1\end{equation}}
\newcommand{\eqs}[1]{\begin{equation}\begin{split}#1\end{split}\end{equation}}

\def\lsim{\mathrel{\hbox{\rlap{\lower.55ex \hbox{$\sim$}} \kern-.3em \raise.4ex \hbox{$<$}}}}
\def\gsim{\mathrel{\hbox{\rlap{\lower.55ex \hbox{$\sim$}} \kern-.3em \raise.4ex \hbox{$>$}}}}

\makeatletter
\def\@fpheader{\ }
\makeatother

\graphicspath{ {Images/} }

\title{Bootstrapping Witten diagrams via differential representation in Mellin space}
\author{Yue-Zhou Li$^{1}$, Jiajie Mei$^2$}
\affiliation{
${}^1$Department of Physics, McGill University, 3600 Rue University, Montr\'eal, H3A 2T8, QC Canada \\
${}^2$Department of Mathematical Sciences, Durham University, Durham, DH1 3LE, United Kingdom\\
}
\emailAdd{yue.z.li@mail.mcgill.ca, jiajie.mei@durham.ac.uk}
\date{}
\abstract{We explore the use of the differential representation of AdS amplitudes to compute Witten diagrams. The differential representation expresses AdS amplitudes in terms of conformal generators acting on contact Witten diagrams, which allows us to construct differential equations for Witten diagrams. These differential equations can then be transformed into difference equations in Mellin space, which can be solved recursively. Using this method, we efficiently re-computed scalar four-point amplitudes and obtained new results for scalar six-point amplitudes mediated by gluons and scalars, as well as two examples of scalar eight-point amplitudes from gluon exchange.}

\begin{document}

\maketitle
\pagenumbering{roman}
\setcounter{page}{2}
\newpage
\pagenumbering{arabic}
\setcounter{page}{1}

\section{Introduction}

The AdS/CFT correspondence \cite{Maldacena:1997re} establishes a conjectural map to identify the gravitational physics in Anti de-Sitter space (AdS) with conformal field theories (CFTs) on the boundary of AdS, providing a potential Ultra-Violet (UV) completion of AdS gravity in terms of the Hilbert space of CFTs \cite{Witten:1998qj}.

The AdS/CFT correspondence has been extensively studied, particularly in the regime where weakly coupled and local gravitational interactions can be described equivalently by CFTs with a large-$N$ limit and a sparse spectrum gap \cite{Heemskerk:2009pn}. In this regime, AdS observables, such as scattering amplitudes, can be evaluated using perturbative techniques in quantum field theory. This is accomplished by generalizing Feynman diagrams to Witten diagrams \cite{Witten:1998qj}. On the CFT side, these amplitudes correspond to the conformal correlation functions in the large-$N$ limit \cite{Freedman:1998tz}. Therefore, the scattering amplitudes in AdS, as represented by Witten diagrams, can shed light on understanding universal structures of large-$N$ CFTs. For example, the four-point amplitude of graviton exchange can be used to build the leading Regge trajectory of large-$N$ CFT \cite{Costa:2012cb}. Witten diagrams can also serve as the expanding blocks of conformal correlators, known as the Polyakov-Regge block, even beyond the large-$N$ limit \cite{Polyakov:1974gs,Mazac:2019shk,Sleight:2019ive}. The study of Polyakov-Regge blocks can benefit the analytic conformal bootstrap program, as seen in works such as \cite{Gopakumar:2016wkt,Caron-Huot:2020adz,Dey:2016mcs,Dey:2017fab,Dey:2017oim,Gopakumar:2018xqi}.

Explicit evaluation of Witten diagrams, even at the tree-level, is notoriously challenging due to the intricate AdS integrals involved. For instance, even the simplest $\phi^4$ contact diagram in position space is expressed by a complex function known as the $D$-function \cite{DHoker:1999kzh,DHoker:2002nbb}. Representing AdS amplitudes through Mellin space is a more natural approach \cite{Penedones:2010ue}. For example, the $D$-function simplifies to a constant in Mellin space. Nevertheless, computing tree-level exchange diagrams in Mellin space using standard techniques is feasible but still remains challenging, particularly for spinning exchange and higher-point amplitudes \cite{Paulos:2011ie,Fitzpatrick:2011ia}.

It is noteworthy that Mellin amplitudes are a natural counterpart to flat-space amplitudes for various reasons. Mellin amplitudes are rational functions of Mellin variables, which are analogous to the flat-space Mandelstam variables from a kinematic perspective. The pole structures of Mellin amplitudes encode the exchanged operators, and factorization applies to the corresponding residues \cite{Goncalves:2014rfa,Fitzpatrick:2011hu}, similar to the analytic structures of flat-space amplitudes. The scaling limit of Mellin variables enables reconstructing the flat-space amplitudes \cite{Penedones:2010ue,Paulos:2016fap}, thus realizing the flat-space limit \cite{Polchinski:1999ry, Giddings:1999jq, Gary:2009ae, Gary:2009mi, Heemskerk:2009pn}. Therefore, it is natural to speculate that there exists a way to efficiently compute Mellin amplitudes by imposing correct analytic structures and consistency with the flat-space limit, i.e., bootstrapping the Mellin amplitudes. Such a strategy turns out to work excellently for supersymmetric correlators, as supersymmetry imposes stronger consistency conditions. See, e.g., \cite{Rastelli:2016nze,Rastelli:2017udc,Rastelli:2017ymc,Zhou:2018ofp,Goncalves:2019znr,Goncalves:2023oyx}. This paper explores how to realize a similar idea for non-supersymmetric amplitudes.

One crucial step in understanding AdS amplitudes is to make the flat-space structures explicit. This can be achieved using the recently developed differential representation \cite{Herderschee:2022ntr,Cheung:2022pdk}. The differential representation expresses amplitudes as differential operators acting on a contact diagram. This approach allows for lifting the flat-space amplitudes to AdS schematically \cite{Eberhardt:2020ewh,Roehrig:2020kck,Diwakar:2021juk,Herderschee:2022ntr,Gomez:2021qfd,Gomez:2021ujt,Cheung:2022pdk,Li:2022tby}. The differential representation in position space yields the Casimir differential equation for  Witten diagrams \cite{DHoker:1999kzh,Binder:2020raz,Zhou:2018sfz,DHoker:1999mqo,Bissi:2022mrs}. Fourier-transforming this differential equation to momentum space yields the bootstrap equation for cosmological correlators presented in \cite{Arkani-Hamed:2018kmz,Baumann:2022jpr} and sheds light on the new structure of correlators in momentum space \cite{Armstrong:2022csc,Armstrong:2022mfr,Armstrong:2022vgl}.

In this paper, we present a novel approach to efficiently compute AdS amplitudes utilizing the differential representation in Mellin space. While our focus is on scalar amplitudes, we allow for spinning particles to be exchanged. Our bootstrapping strategy, as previously highlighted, combines the correct analytic structures of Mellin amplitudes with the differential representation. Specifically, given a Witten diagram, we provide a general ansatz for the Mellin amplitudes with poles encoding single-trace operators. The differential representation constructs difference equations, enabling us to determine the coefficients in our ansatz and consequently solve for AdS amplitudes efficiently. This approach allows us to quickly reproduce four-point scalar amplitudes with scalar, gluon, and graviton exchange, for any spacetime and scaling dimension. Using our bootstrap method, we also compute higher-point amplitudes, such as six-point scalar amplitudes with gluon exchange, and two examples of eight-point amplitudes. These amplitudes were previously challenging to compute using standard techniques \cite{Paulos:2011ie,Fitzpatrick:2011ia}.

This paper is organized as follows. In Section \ref{sec: general}, we begin by providing a brief review of the embedding formalism, Witten diagrams, and the definition of Mellin amplitudes. We then explain how the Casimir equations for Witten diagrams remove the single-trace operator contribution. In Section \ref{sec:Algorithm}, we outline an algorithm for computing AdS amplitudes in Mellin space and present several examples of four-point amplitudes. In Section \ref{sec:Higher}, we extend the algorithm to higher-point Mellin amplitudes and include several examples such as a six-point snowflake with scalar exchanges, a six-point snowflake with gluon exchanges, and several truncated six-point amplitudes. We also provide two simple examples of eight-point amplitudes. In Section \ref{sec: remarks}, we offer further remarks on the relation between our work and cosmological correlators as well as the flat-space limit. Finally, in Appendix \ref{sec: rules1}, we explicitly derive the boundary Feynman rules involved in graviton exchange and summarize the Feynman rules for future reference. In Appendix \ref{sec: rules2}, we review the scalar Mellin Feynman rules.

\section{Generalities}
\label{sec: general}

\subsection{The embedding formalism}

In this paper, we work with the embedding formalism of Euclidean AdS, see \cite{Costa:2014kfa} for more details. We use $Y$ to denote the coordinates of $d+2$ dimensional Minkowski space $\mathbb{M}^{d+2}$, which is referred to as the embedding space. and AdS$_{d+1}$ is defined by a hyperboloid subjecting to
\be
Y^2=-R_{\rm AdS}^2\,.
\ee 
We take $R_{\rm AdS}=1$ without further notice. It is instructive to think of Euclidean AdS$_{d+1}$ as a coset space ${\rm SO}(d+1,1)/{\rm SO}(d+1)$, so as to apply harmonic analysis and convert to the isometric frame for analyzing kinematics \cite{Cheung:2022pdk}. Fields of AdS$_{d+1}$ can be classified by finite-dimensional irrep of ${\rm SO}(d+1)$ in the embedding space, transverse to the hyperboloid. We are mostly interested in the spin-$J$ traceless symmetric tensors and it benefits to represent them as index-free polynomials in embedding polarizations $W$
\be
F(Y,W)= W^{A_1}\cdots W^{A_J}F_{A_1\cdots A_J}\,,
\ee
subject to
\be
Y\cdot W=W^2=0\,.
\ee 
It is not hard to generalize to mixed-symmetric tensors by using different $W$ for different rows in Young diagram \cite{Costa:2018mcg}; see also \cite{Binder:2020raz} for the generalization to spinor fields. By this construction, the contractions between two tensors in the index-free notation can be written by summing over polarizations, namely
\be
F_{A_1\cdots A_J}H^{A_1\cdots A_J} = \sum_{W} F(W^\ast)H(W)=\fft{1}{J!(\fft{d-1}{2})_J}F(K^W)H(W)\,,
\ee
where the differential operator $K_A$ is to restore the indices from polynomials \cite{Costa:2014kfa} 
\be
K_A^W = \big(\fft{\partial}{\partial W^A}+Y_A (Y\cdot\fft{\partial}{\partial W})\big)\big(\fft{d-3}{2}+W\cdot\fft{\partial}{\partial W}\big) -\fft{1}{2}W_A \big(\fft{\partial^2}{\partial W^2} + (Y\cdot\fft{\partial}{\partial W})^2\big)\,.
\ee
We also need the covariant AdS derivatives acting on symmetric traceless tensors encoded in polynomials of $W$ \cite{Costa:2014kfa} 
\be
\nabla_A^Y = \fft{\partial}{\partial Y} + Y_A \big(Y\cdot \fft{\partial}{\partial Y}\big) + W_A\big(Y\cdot \fft{\partial}{\partial W}\big)\,.
\ee

The boundary of AdS can also be embedded into $\mathbb{M}^{d+1}$ as a lightcone surface, known as the embedding formalism for CFT \cite{Costa:2011mg}, where we denote the coordinate by $X$ and polarization as $Z$ subject to
\be
X^2=X\cdot Z=Z^2=0\,,
\ee
with ``gauge'' redundancy $Z\simeq Z+\# X$.

Using the embedding formalism, the spin-$J$ symmetric traceless bulk-to-boundary propagator $\Pi_{\Delta,J}(Y,W;X,Z)$ with conformal dimension $\Delta$ satisfies the equation
\be
\big((\nabla^Y)^2 -\Delta(\Delta-d)+J\big) \Pi_{\Delta,J}(Y,W;X,Z)  = 0\,,
\ee
which is solved by
\be
\Pi_{\Delta,J}(Y,W;X,Z)= \mathcal{C}(\Delta,J) \fft{\big(2(W\cdot X)(Z\cdot Y)-2 (X\cdot Y)(W\cdot Z)\big)^J}{(-2X\cdot Y)^{\Delta+J}}\,,
\ee
where
\be
\mathcal{C}(\Delta,J) = \fft{(J+\Delta-1)\Gamma(\Delta)}{2\pi^{\fft{d}{2}}(\Delta-1)\Gamma(\Delta+1-\fft{d}{2})}\,.
\ee
In this paper, we only study scattering between external scalars, where only $\Pi_{\Delta,0}$ and its derivatives are involved.

\subsection{Witten diagrams and the bulk Feynman rules}
\label{subsec: bulk rules}
In this section, we review the derivation of Witten diagrams from wick contraction in AdS/CFT for external scalars. The essential formula is the holographic dictionary \cite{Witten:1998qj}
\be
Z_{\rm bulk} = \int [D\Phi]e^{-S[\Phi]} = \langle e^{\int \sum_i\phi_i^{(0)} \mathcal{O}_i}\rangle_{\rm CFT}\,,
\ee
where $\Phi$ denotes all possible fields in the bulk, $\phi^{(0)}_i$ is the boundary value of external scalar $\phi_i$ whereas $\mathcal{O}_i$ is the dual primary operator. The CFT correlator can then be evaluated by
\be
\langle\mathcal{O}_1\cdots\mathcal{O}_n\rangle= \Big(\prod_i \fft{\delta}{\delta\phi_i^{(0)}} \Big) \langle e^{\int \sum_i\phi_i^{(0)} \mathcal{O}_i}\rangle_{\rm CFT}=
\Big(\prod_i \fft{\delta}{\delta\phi_i^{(0)}}\Big) \langle e^{-S_{\rm int}[\Phi]}\rangle_{\rm bulk}\,,
\ee
where the bulk expectation value is performed for free theory. The functional derivatives $\delta/\delta\phi^{(0)}$ essentially replace bulk field $\phi$ by the corresponding bulk-to-boundary propagator because
\be
\phi_i(Y) = \int d^{d+2}X \Pi_{\Delta_i}(Y,X) \phi^{(0)}_i(X)\,.\ee
Then we can work on the wick contraction for the rest fields. The wick contraction pairly groups the same (or complex conjugate, depending on the representation of fields) bulk field contents under the bulk expectation value, giving rise to products of bulk-to-bulk propagators (probably with bulk derivatives).  This procedure generates bulk Feynman rules. 

\begin{figure}[h]
\centering
 \begin{tikzpicture}
        \draw (0,0) circle (2 cm);
        \draw (1,0) circle (0.5 cm);
        \coordinate (1) at (-1.41421,1.41421);
        \coordinate (2) at (-1.84776,0.765367);
        \coordinate (n) at (-1.41421,-1.41421);
        \coordinate (C) at (-0.75,0);
        \coordinate (Phi) at (0.5,0);
        \coordinate (4) at (1.84776,-0.765367);
        \coordinate (3) at (1.84776,0.765367);
        \coordinate (5) at (1.353553,0.353553);
        \coordinate (6) at (1.353553,-0.353553);

        \node at  (-1.15899, -0.0445895) {$\cdots$};
        \node at  (1.7,0.2) {$\cdot$};
        \node at  (1.7,-0.2) {$\cdot$};
        
        \draw (1) -- (C);
        \draw (2) -- (C);
        \draw (n) -- (C);
        \draw (3) -- (5);
        \draw (4) -- (6);
        \draw[line width=1.5pt] (C) -- (Phi);
        
        \begin{scope}
            \clip (0,0) circle (2 cm);
            \fill[pattern=north east lines](1,0) circle (0.5 cm);
        \end{scope}

        \fill (1)  node[left] {$1$};
        \fill (2) node[left] {$2$};
        \fill (n) node[left] {$n$};
        \fill (C) circle (2pt) node[above right] {$\Phi$};
        \fill (Phi);
    \end{tikzpicture}
\caption{A general $n+1$-point vertex in a Witten diagram, where there is a bulk-to-bulk propagator with any possible spin associated with the field $\Phi$, represented by the bold line. The shaded region on the right denotes any possible tree-level processes that are irrelevant to the ongoing discussion.}
\label{fig: n+1 vert}
\end{figure}
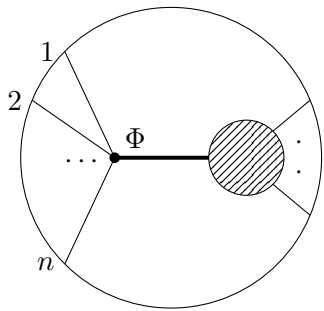

A single Witten diagram, although not crossing symmetric, still admits the operator product expansion (OPE) because of the conformal symmetry. A salient property of Witten diagram is that the bulk-to-bulk propagator with conformal dimension $\Delta$ and spin $J$ signals the existence of single-trace primary operator $\mathcal{O}_{\Delta,J}$ in the OPE. To observe this fact, we study a general $n+1$ point vertex with one bulk-to-bulk propagator, as shown in Fig \ref{fig: n+1 vert}. 

This vertex, as a part of a Witten diagram, can be schematically written down after integration-by-parts. One example is
\be
\int D^{d+2}Y \big(\prod_{i=1}^n(\nabla^Y\cdots)_{\alpha_{ij}}(W^\ast\cdot\nabla^Y)^J\Pi_{\Delta_i}(Y,X_i)\big)\Big\langle \Phi(Y,W)\cdots \Big\rangle\,,
\ee
where $(\nabla^Y\cdots)_{\alpha_{ij}}$ denotes that there are $\alpha_{ij}$ covariant derivatives are contracted between $i$-th and $j$-th bulk-to-boundary propagators, and $\Phi$ is a spin-$J$ symmetric traceless tensor field. We can also change the ordering of derivatives to enumerate all possible vertices. Without losing generality, any such term, as derivatives on bulk-to-boundary propagators, must be the sum of functions of $X_i\cdot X_j, X_i\cdot W^\ast$ and $X_i\cdot Y$ in the following format
\be
P^{\Delta_i}_{a_i,b_i,\beta_{ij}}=\prod_{i=1}^n \fft{(X_i\cdot W^\ast)^{a_i-b_i}}{(X_i\cdot Y)^{\Delta_i+a_i}}\prod_{i<j}(X_i\cdot X_j)^{\beta_{ij}}\,,\quad b_i=\sum_{j\neq i}\beta_{ij}\,,
\ee
This function satisfies the following identity involving the Casimir operator and AdS Laplacian
\bea
\mathcal{C}_{1\cdots n}P^{\Delta_i}_{a_i,b_i,\beta_{ij}} = \Big((\nabla^Y)^2+J(J+d-1)\Big)P^{\Delta_i}_{a_i,b_i,\beta_{ij}} \,,\label{eq: identity Casimir}
\eea
where the quadratic Casimir operator involving $n$-points $\mathcal{C}_{1\cdots n}$ is constructed by conformal generators $L_i^{AB}$
\be
\mathcal{C}_{1\cdots n}=\fft{1}{2} \big(\sum_{i=1}^n L_i^{AB}\big)^2\,,\quad L_i^{AB}=-i\Big(X_i^A \fft{\partial}{\partial X_{iB}}-X_i^B \fft{\partial}{\partial X_{iA}}\Big)\,.
\ee
Using the identity \eqref{eq: identity Casimir}, we can see that the differential operator
\be
\mathcal{D}_{1\cdots n}^{\Delta,J}=\mathcal{C}_{1\cdots n} -C(\Delta,J)\,,\quad C(\Delta,J)=\Delta(\Delta-d)+J(J+d-2)\,,
\ee
as acting on the Witten diagram, eliminates the bulk-to-bulk propagator in the corresponding vertex. Because we can move the resulting differential operator $(\nabla^Y)^2-\Delta(\Delta-d)+J$ from $P_{a_i,b_i,\beta_{ij}}^{\Delta_i}$ to the relevant bulk-to-bulk propagator by integration-by-parts and recall the equation of motion \cite{Costa:2014kfa} 
\be
\Big((\nabla^Y)^2-\Delta(\Delta-d)+J\Big)\Pi_{bb}^{\Delta,J}(Y,Y^\prime;W,W^\prime) = -\delta^{(d+2)} (Y-Y^\prime)(W_1\cdot W_2)^{J}\,.\label{eq: traceless eq}
\ee
This is saying the transverse and traceless bulk-to-bulk propagator is removed by Casimir equation with quantum number $(\Delta,J)$, proving it contributes single-trace operator in the OPE. Nevertheless, this is not the end of story. Spinning fields can also contribute the longitudinal and trace part when two bulk points they connect are the same, namely the longitudinal and trace part contribute nontrivial contact terms via exchanging the longitudinal and trace modes, respectively. It is rather easy to deal with this problem for gauge fields (i.e., conserved operators on the boundary) such as gluon and graviton. In this case, the longitudinal modes are nonphysical and we can take the convenient gauge choice to simplify the calculations. In this paper, we take the Feynman gauge for gluon, and the de Donder gauge for graviton. However, it is worth noting that some Witten diagram has to be combined with others to ensure gauge invariance (or at least to ensure consistency with our gauge choice). Such diagram requires cautious treatments, as we will show in section \ref{subsec: subtle non-gauge}. On the other hand, to deal with the trace modes, we can decompose the fields into traceless part and trace part so that we have
\be
\Big\langle {\rm Tr}\,\Phi(Y,W)\cdots\Big\rangle\,,
\ee
in the bulk Feynman rules. Notice ${\rm Tr}\,\Phi(Y,W)$ has two less $W$s and thus carries spin $J-2$. This trace term is also removed by $\mathcal{D}_{1\cdots}^{\Delta,J}$, because it simply shifts \eqref{eq: identity Casimir} by $J\rightarrow J-2$ and consequently we end up with precisely the relevant equation \cite{Sleight:2016hyl}
\be
\Big((\nabla^Y)^2 - J(J+d-1) +2\Big)({\rm Tr}\,\Pi)_{bb}^{\Delta,J}(Y,Y^\prime;W,W^\prime) = 
\fft{J(J-1) (W_1\cdot W_2)^{J-2}\delta^{(d+2)} (Y-Y^\prime)}{(d+2J-3)(d+2J-5)}\,.\label{eq: trace eq}
\ee
We implement such example for graviton in detail in appendix \ref{subsec: gravity rules}. 

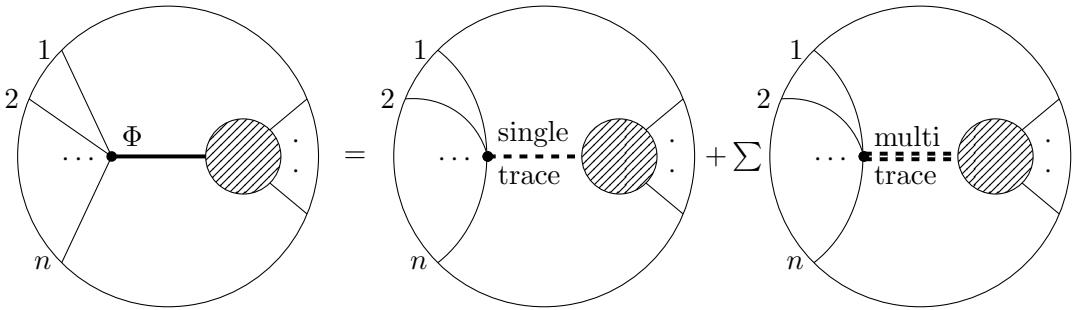
\begin{figure}[h]
\centering
 \begin{tikzpicture}
        \draw (0,0) circle (2 cm);
        \draw (1,0) circle (0.5 cm);
        \coordinate (1) at (-1.41421,1.41421);
        \coordinate (2) at (-1.84776,0.765367);
        \coordinate (n) at (-1.41421,-1.41421);
        \coordinate (C) at (-0.75,0);
        \coordinate (Phi) at (0.5,0);
        \coordinate (4) at (1.84776,-0.765367);
        \coordinate (3) at (1.84776,0.765367);
        \coordinate (5) at (1.353553,0.353553);
        \coordinate (6) at (1.353553,-0.353553);

        \node at  (-1.15899, -0.0445895) {$\cdots$};
        \node at  (1.7,0.2) {$\cdot$};
        \node at  (1.7,-0.2) {$\cdot$};
        
        \draw (1) -- (C);
        \draw (2) -- (C);
        \draw (n) -- (C);
        \draw (3) -- (5);
        \draw (4) -- (6);
        \draw[line width=1.5pt] (C) -- (Phi);
        
        \begin{scope}
            \clip (0,0) circle (2 cm);
            \fill[pattern=north east lines](1,0) circle (0.5 cm);
        \end{scope}

        \fill (1)  node[left] {$1$};
        \fill (2) node[left] {$2$};
        \fill (n) node[left] {$n$};
        \fill (C) circle (2pt) node[above right] {$\Phi$};
        \fill (Phi);
        \node[anchor=west, right=1.7 cm of Phi] (formula) {\(= \)};  
        \draw (5,0) circle (2 cm);
        \draw (6,0) circle (0.5 cm);
        
        \coordinate (1p) at (3.58579,1.41421);
        \coordinate (2p) at (3.15224,0.765367);
        \coordinate (np) at (3.58579,-1.41421);
        \coordinate (Cp) at (4.25,0);
        \coordinate (Phip) at (5.5,0);
        \coordinate (4p) at (6.84776,-0.765367);
        \coordinate (3p) at (6.84776,0.765367);
        \coordinate (5p) at (6.353553,0.353553);
        \coordinate (6p) at (6.353553,-0.353553);
        
        \node at (3.84101,-0.0445895) {$\cdots$};
         \node at  (6.7,0.2) {$\cdot$};
        \node at  (6.7,-0.2) {$\cdot$};
       \draw (3p) -- (5p);
        \draw (4p) -- (6p);
        \draw[dashed, line width=1.5pt] (Cp) -- (Phip);
        
        \draw[] (1p) to [bend left=52] (3.58579,-1.41421);
         \draw[] (2p) to [bend left=40] (4.25,0);
        
        \begin{scope}
            \clip (5,0) circle (2 cm);
            \fill[pattern=north east lines](6,0) circle (0.5 cm);
        \end{scope}

        \fill (1p)  node[left] {$1$};
        \fill (2p) node[left] {$2$};
        \fill (np) node[left] {$n$};
        \fill (Cp) circle (2pt) node[above right] {single};
        \fill (Cp) circle (2pt) node[below right] {trace};
        \fill (Phip);
         \node[anchor=west, right=1.5 cm of Phip] (formula) {\(+ \sum\)};  
        \draw (10,0) circle (2 cm);
         \draw (11,0) circle (0.5 cm);
        
        \coordinate (1pp) at (8.58579,1.41421);
        \coordinate (2pp) at (8.15224,0.765367);
        \coordinate (npp) at (8.58579,-1.41421);
        \coordinate (Cpp) at (9.25,0);
        \coordinate (Phipp) at (10.5,0);
         \coordinate (4pp) at (11.84776,-0.765367);
        \coordinate (3pp) at (11.84776,0.765367);
        \coordinate (5pp) at (11.353553,0.353553);
        \coordinate (6pp) at (11.353553,-0.353553);
        
        \node at (8.84101,-0.0445895) {$\cdots$};
        \node at  (11.7,0.2) {$\cdot$};
        \node at  (11.7,-0.2) {$\cdot$};
        
        \draw (3pp) -- (5pp);
        \draw (4pp) -- (6pp);
        \draw[double, dashed, line width=1.5pt] (Cpp) -- (Phipp);
        
        \draw[] (1pp) to [bend left=52] (8.58579,-1.41421);
         \draw[] (2pp) to [bend left=40] (9.25,0);
        
        \begin{scope}
            \clip (10,0) circle (2 cm);
            \fill[pattern=north east lines](11,0) circle (0.5 cm);
        \end{scope}

        \fill (1pp)  node[left] {$1$};
        \fill (2pp) node[left] {$2$};
        \fill (npp) node[left] {$n$};
        \fill (Cpp) circle (2pt) node[above right] {multi};
         \fill (Cpp) circle (2pt) node[below right] {trace};
        \fill (Phipp);
    \end{tikzpicture}
\caption{An $n+1$ point vertex with a bulk-to-bulk propagator $\Phi$ can be decomposed into a single-trace operator and a sum of multi-trace operators. We use the curved line to emphasize that it is NOT a Witten diagram, but rather a decomposition into conformal blocks represented by using the curved lines. The dashed line refers to the exchanged single-trace operator, and the double dashed line refers to the exchanged multi-trace operators.}
\label{fig: decompose Witten}
\end{figure}

For contact diagram, there are no single-trace exchanges and the OPEs only contain multi-trace operators \cite{DHoker:1999kzh}. For any vertex with one bulk-to-bulk propagator, we can decompose it into a single-trace operator and the sum over multi-trace operators, see Fig \ref{fig: decompose Witten}.

\subsection{Casimir equations for Witten diagrams}

\begin{figure}[h]
\centering
 \begin{tikzpicture}
        \draw (0,0) circle (2 cm);
        
        \coordinate (1) at (-1.41421,1.41421);
        \coordinate (2) at (-1.41421,-1.41421);
        \coordinate (C) at (-1,0);
        \coordinate (D) at (1,0);
         \coordinate (4) at (1.41421,1.41421);
        \coordinate (3) at (1.41421,-1.41421);
              
        \draw (1) -- (C);
        \draw (2) -- (C);
        \draw (C) -- (D) node[midway, above] {\((\Delta,J)\)};
         \draw (3) -- (D);
        \draw (4) -- (D);
                  
        \fill (1)  node[above] {$1$};
        \fill (2) node[below] {$2$};
        \fill (3) node[below] {$3$};
         \fill (4) node[above] {$4$};
       \end{tikzpicture}
\caption{The general four-point Witten diagram, where the exchanged single-trace operator has the scaling dimension $\Delta$ and spin $J$.}
\label{fig: 4pt Witten}
\end{figure}
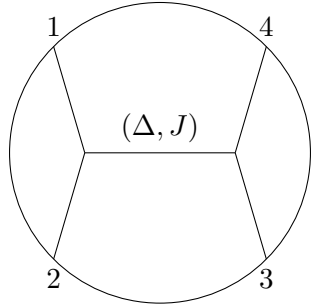

\begin{figure}[h]
\centering
 \begin{tikzpicture}
        \draw (0,0) circle (2 cm);
          \draw (1,0) circle (0.5 cm);
        \coordinate (1) at (-1.41421,1.41421);
        \coordinate (2) at (-1.84776,0.765367);
        \coordinate (n) at (-1.41421,-1.41421);
        \coordinate (C) at (-0.75,0);
        \coordinate (Phi) at (0.5,0);
        \coordinate (4) at (1.84776,-0.765367);
        \coordinate (3) at (1.84776,0.765367);
        \coordinate (5) at (1.353553,0.353553);
        \coordinate (6) at (1.353553,-0.353553);

        \node at (-1.15899, -0.0445895) {$\cdots$};
         \node at  (1.7,0.2) {$\cdot$};
        \node at  (1.7,-0.2) {$\cdot$};
        
        \draw (1) -- (C);
        \draw (2) -- (C);
        \draw (n) -- (C);
        \draw (3) -- (5);
        \draw (4) -- (6);
        \draw[line width=1.5pt] (C) -- (Phi);
        
        \begin{scope}
            \clip (0,0) circle (2 cm);
            \fill[pattern=north east lines](1,0) circle (0.5 cm);
        \end{scope}
              
        \fill (1)  node[left] {$1$};
        \fill (2) node[left] {$2$};
        \fill (n) node[left] {$n$};
        \fill (C) node[midway, above] {$(\Delta,J)$};
        \fill (Phi);
        \node[anchor=west, left =2.8 cm of Phi] (formula) {\(\mathcal{D}_{12\cdots n}^{\Delta,J}\)};  
        \node[anchor=west, right=1.5 cm of Phi] (formula) {\(\,:= \)};  
        
        \draw (5,0) circle (2 cm);
        \draw (6,0) circle (0.5 cm);
        
        \coordinate (1p) at (3.58579,1.41421);
        \coordinate (2p) at (3.15224,0.765367);
        \coordinate (np) at (3.58579,-1.41421);
        \coordinate (Cp) at (4.25,0);
        \coordinate (Phip) at (5.5,0);
        \coordinate (4p) at (6.84776,-0.765367);
        \coordinate (3p) at (6.84776,0.765367);
        \coordinate (5p) at (6.353553,0.353553);
        \coordinate (6p) at (6.353553,-0.353553);
         \coordinate (cutup) at (5,2.4);
        \coordinate (cutdown) at (5,-2.4);
        
        \node at (3.84101,-0.0445895) {$\cdots$};
         \node at  (6.7,0.2) {$\cdot$};
        \node at  (6.7,-0.2) {$\cdot$};
       \draw (3p) -- (5p);
        \draw (4p) -- (6p);
        \draw (1p) -- (Cp);
        \draw (2p) -- (Cp);
        \draw (np) -- (Cp);
        \draw (Cp) -- (Phip) node[midway, above] {$(\Delta,J)$};
        \draw[dashed] (cutup) -- (cutdown);
        
        \begin{scope}
            \clip (5,0) circle (2 cm);
            \fill[pattern=north east lines](6,0) circle (0.5 cm);
        \end{scope}

        \fill (1p)  node[left] {$1$};
        \fill (2p) node[left] {$2$};
        \fill (np) node[left] {$n$};
        \fill (Cp);
        \fill (Phip);
        \node[anchor=west, right=1.6 cm of Phip] (formula) {\(\,=\)};  

        \draw (10,0) circle (2 cm);
         \draw (11,0) circle (0.5 cm);
        
        \coordinate (1pp) at (8.58579,1.41421);
        \coordinate (2pp) at (8.15224,0.765367);
        \coordinate (npp) at (8.58579,-1.41421);
        \coordinate (Cpp) at (9.25,0);
        \coordinate (Phipp) at (10.5,0);
         \coordinate (4pp) at (11.84776,-0.765367);
        \coordinate (3pp) at (11.84776,0.765367);
        \coordinate (5pp) at (11.353553,0.353553);
        \coordinate (6pp) at (11.353553,-0.353553);

        \node at (9.84101,-0.0445895) {$\cdots$};
        \node at  (11.7,0.2) {$\cdot$};
        \node at  (11.7,-0.2) {$\cdot$};

        \draw (1pp) -- (Phipp);
        \draw (2pp) -- (Phipp);
        \draw (npp) -- (Phipp);
        \draw (3pp) -- (5pp);
        \draw (4pp) -- (6pp);
        
        \begin{scope}
            \clip (10,0) circle (2 cm);
            \fill[pattern=north east lines](11,0) circle (0.5 cm);
        \end{scope}
              
        \fill (1pp)  node[left] {$1$};
        \fill (2pp) node[left] {$2$};
        \fill (npp) node[left] {$n$};
        \fill (Phipp) circle (2pt);
        \end{tikzpicture}
\caption{The action of the differential operator $\mathcal{D}_{12\cdots n}^{\Delta,J}$ is visualized as a Casimir cut, denoted by the vertical dashed line. The Casimir cut removes the corresponding bulk-to-bulk propagator and results in a new Witten diagram with a dotted effective vertex.}
\label{fig: Casimir cut}
\end{figure}
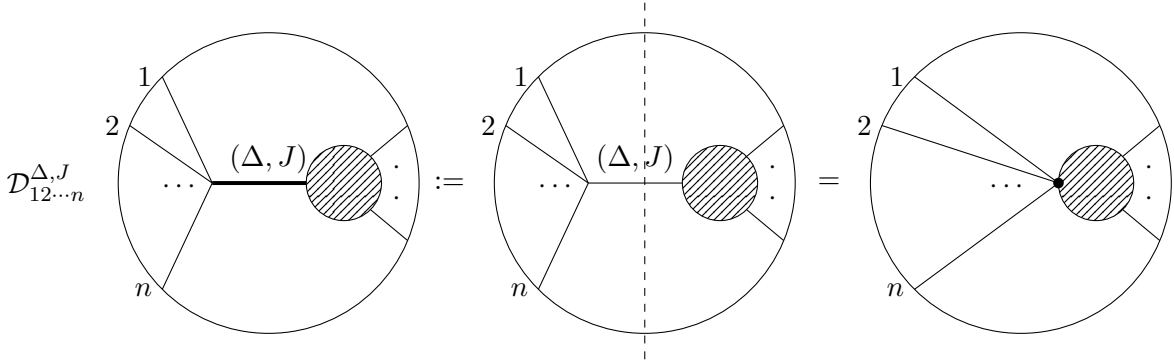

Since the Casimir operator subtracting the corresponding eigenvalue can remove the bulk-to-bulk propagator, we then have Casimir equations for determining a single Witten diagram. The simplest example is four-point amplitude, as shown in Fig \ref{fig: 4pt Witten}. 
$\mathcal{D}_{12}^{\Delta,J}$ effectively modifies the OPE by prefactor but essentially keeps conformal blocks intact
\be
C(\Delta_1+\Delta_2+2n+J^\prime,J^\prime) - C(\Delta,J)\,.
 \ee
As a result, we arrive at an important identity
\be
\mathcal{D}_{12}W_{\Delta,J}^{1234} = \tilde{W}^{1234}_{\rm contacts}\,,\label{eq: Witten4pt}
\ee
where $\tilde{W}$ denotes the resulting contract contributions from acting with $\mathcal{D}_{12}$. Eq.~\eqref{eq: Witten4pt} and its generalization to higher-points (see eq.~\eqref{eq: cutting} below) are essential equations for us to compute Witten diagrams. The explicit information of vertices now enter $\tilde{W}^{1234}_{\rm contacts}$ and we expect there are corresponding Feynman rules to determine what is $\tilde{W}^{1234}_{\rm contacts}$ for given the bulk vertices. As noticed in \cite{Cheung:2022pdk,Herderschee:2022ntr} it is instructive to define the inverse operator $\mathcal{D}_{12}^{-1}$ and then any tree-level exchanged diagram can be written as $\mathcal{D}_{12}^{-1}\times (\text{Feynman rules for vertices})$, which resembles the flat-space structure $1/s \times (\text{Feynman rules for vertices})$.
Similar for higher-point Witten diagrams, we can insert the Casimir equation operator for each propagator that connects $n$-point external legs via $n+1$-point vertex to obtain a new Witten diagram with one less propagator
\be
\mathcal{D}_{12\cdots n}^{\Delta,J} W^{1\cdots n\cdots}_{V,P} = W^{1\cdots n\cdots}_{V-1,P-1}\,.\label{eq: cutting}
\ee
where $V$ and $P$ denote the number of vertices and propagators respectively. This insertion can still be visualized as a cut that we call the Casimir cut, see Fig \ref{fig: Casimir cut} for an illustration. 

The Casimir cut is on the right opposite of the unitarity cut (i.e., the double discontinuity \cite{Caron-Huot:2017vep,Simmons-Duffin:2017nub,Meltzer:2019nbs}): the unitarity cut keeps only the single-trace while the Casimir cut kills the single-trace. By recursively applying the Casimir cut \eqref{eq: cutting} , we can arrive at pure contact diagrams which are determined by underlying Feynman rules.

However, this is not the end of story. It would be helpful to reformulate the complicated contact derived from Casimir cuts as boundary differential operators that act on known functions, e.g., $D_{\Delta_i}$ function constructed by $\phi^n$ contact. The simplest such differential operator is the conformal generator. We refer to the rules of writing contacts in terms of conformal generators acting on $D_{\Delta_i}$ as the boundary Feynman rules. To achieve this, we recall that AdS$_{d+1}$ is the coset space ${\rm SO}(d+1,1)/{\rm SO}(d+1)$, and thus employ the isometric frame techniques developed in \cite{Cheung:2022pdk} to rewrite the contact diagrams. Refer to \cite{Cheung:2022pdk} for more details. The idea is to use the Killing vectors
\be
\mathcal{K}_{AB}=\mathcal{K}_{AB}\,^C \partial^Y_C\,,\quad \mathcal{K}_{AB}\,^C=Y_A \delta_B^C -Y_B \delta_A^C\,,
\ee
to rewrite contractions by replacing the AdS metric
\be
g_{AB}=\eta_{AB}+Y_A Y_B = g_{CDEF}\mathcal{K}^{CD}\,_{A}\mathcal{K}^{EF}\,_{B}\,
\ee
where $g_{CDEF}$ is the Killing metric
\be
g_{CDEF}=\fft{1}{4}(\eta_{CE}\eta_{DF}-\eta_{CF}\eta_{DE})\,.
\ee
The Killing vectors can push the covariant derivatives to be the bulk Lorentz generator and further to be the conformal generator by using isometry, namely
\be
\mathcal{K}_{BC}\,^A \nabla_A^Y \Pi_{\Delta}(Y,X) = -i L_{BC} \Pi_{\Delta}(Y,X)\,.\label{eq: from derivative to generator}
\ee
If we are considering two-derivative theories, where any bulk-to-boundary propagator has at most one derivative,  the boundary Feynman rules are straightforward using \eqref{eq: from derivative to generator}. This is the case that we mostly encounter in this paper. 

More generally, for higher derivatives, we can still move the Killing vectors into derivatives to transform a particular derivative to the conformal generator, trading by additional terms involving derivatives on Killing vectors. The derivatives of Killing vectors are also contracting with Killing vectors, which can be expanded by multiple Killing vectors together with structure constant $f_{ABCD}\,^{EF} = 32 g_{ABG}\,^H g_{CDH}\,^I g^{EF}\,_I\,^G$, modulo $Y^2=-1$ due to the completeness\footnote{This statement is not rigorous because we cannot prove it. Nevertheless, supporting examples demonstrate that any remaining bulk coordinates $Y$ and AdS metric $g_{AB}$ can be eliminated by employing specific bulk identities. Refer to \cite{Li:2022tby} for the case involving gluons and gravitons.}, giving rise to terms with only conformal generator (but less). The simplest such example is 
\be
\mathcal{K}_{AB}\,^C \nabla^Y_C \mathcal{K}_{EF}\,^G -\mathcal{K}_{EF}\,^C \nabla^Y_C \mathcal{K}_{AB}\,^G=  f_{ABEF}\,^{CD} \mathcal{K}_{CD}\,^G\,.
\ee
We have not found a systematic way to perform this analysis, but it is not hard to work out simple examples. For example, we find
\be
&\int D^{d+2}Y \big(\nabla^Y_A \nabla^Y_B \Pi_{\Delta_1}(Y,X_1)\big)\big(\nabla^{YA} \nabla^{YB} \Pi_{\Delta_2}(Y,X_2)\big) \Pi_{\Delta_3}(Y,X_3)\Pi_{\Delta_4}(Y,X_4)=\nn\\
& \Big(\fft{1}{4} (L_1\cdot L_2)^2 + \fft{d-1}{2} L_1\cdot L_2\Big)D_{\Delta_1\Delta_2\Delta_3\Delta_4}\,.
\ee
More examples can be found in \cite{Herderschee:2022ntr,Armstrong:2022csc}.

After having the boundary Feynman rules, we then have Casimir equations for any single $n$-point Witten diagram, stating that Casimir equation operators acting on certain Witten diagrams equals the conformal generators acting on the $D$-function.

\subsection{Mellin amplitudes}

Although we can translate any Witten diagram to solving differential equations, the differential equation is still highly nontrivial and intricate to solve. The Mellin representation of correlator would make life much easier and is also natural to check the flat-space limit \cite{Penedones:2010ue}
\be
\langle\mathcal{O}_1\cdots \mathcal{O}_n\rangle = \fft{\mathcal{N}}{(2\pi i)^{\fft{n(n-3)}{2}}}\int [d\delta_{ij}]\prod_{i<j}\fft{\Gamma(\delta_{ij})}{(-2X_i\cdot X_j)^{\delta_{ij}}}M(\delta_{ij})\,,\quad \sum_{j\neq i}\delta_{ij}=\Delta_i\,.
\ee
Our convention of normalization is
\be
\mathcal{N}=\fft{\pi^{\fft{d}{2}}}{2}\Gamma\Big(\fft{\Delta_{\Sigma}-d}{2}\Big)\prod_{i=1}^n \fft{\mathcal{C}(\Delta_i,0)}{\Gamma(\Delta_i)}\,,
\ee
where $\Delta_\Sigma=\sum_{i=1}^n \Delta_i$. The Mellin amplitudes $M(\delta_{ij})$  is the object we calculate in this paper. Notice that the integral contour is along the imaginary axis of independent $\delta_{ij}$. By deforming the contour to real axis and picking up poles in Mellin amplitudes and the integral kernel (i.e., Gamma functions), one can get the correlation functions in the coordinate space back. 

Without really doing the integral of Mellin variables, we can still analyze the OPE from Mellin amplitudes by looking into the position of poles for Mellin variables. To analyze the OPE, we can scale the relevant singular distance by $\lambda$ and count the power of $\lambda$. The power of $\lambda$ is a linear combination of $\delta_{ij}$, thus we are able to read off the spectra involved in the OPE from the position of poles in $\delta_{ij}$. From this analysis, one can immediately conclude that if a Mellin amplitude is polynomials in $\delta_{ij}$, it must represent a certain contact diagram in AdS, because the only poles are integers $\delta_{ij}\leq 0$ and those poles are corresponding multi-twist operators. To see this, we can analyze $\mathcal{O}_1\mathcal{O}_2\rightarrow \mathcal{O}$, which gives rise to $\lambda^{-\delta_{ij}}\sim \lambda^{n}$ for integer $n$, perfectly matching with OPE singularity $(x_{12}^2)^{\fft{\Delta-J-(\Delta_1+\Delta_2)}{2}}$ for double-twist operators $\Delta=\Delta_1+\Delta_2+J+2n$. Similarly, for higher point functions, we can also analyze $\mathcal{O}_1\mathcal{O}_2\mathcal{O}_3\rightarrow \mathcal{O}$, finding triple-twist operators. 

On the other hand, whenever there are additional poles in $M(\delta_{ij})$, we conclude there are corresponding single-trace exchanges. For example
\be
M(\delta_{ij}) \sim \fft{\lambda_{12\mathcal{O}}}{\delta_{12}-\fft{\Delta_1+\Delta_2-(\Delta-J)}{2}}\,.
\ee
This pole captures the leading term of single-trace operator $\mathcal{O}_{\Delta,J}$ in $\mathcal{O}_1\mathcal{O}_2$ OPE, where $\lambda_{12\mathcal{O}}$ is the OPE coefficient. There are also subsequent poles in the OPE, more precisely, one has
\be
M(\delta_{ij}) \sim \sum_{n=0,1\cdots}\fft{\lambda_{12\mathcal{O}}f_m}{\delta_{12}-\fft{\Delta_1+\Delta_2-(\Delta-J+2n)}{2}}\,,
\ee
where $f_m$ is function of $\delta_{ij}$ which may also contain poles giving other single-trace operators. In general, Mellin amplitudes must have such infinite sum of pole structures to fully encode single-trace operator, but for special cases where the position of the pole coincides with double-twist operators, the sum truncates there, i.e., 
\be
M(\delta_{ij}) \sim \sum_{n=0}^{n_0-1}\fft{\lambda_{12\mathcal{O}}f_m}{\delta_{12}-\fft{\Delta_1+\Delta_2-(\Delta-J+2n)}{2}}\,,\quad \Delta_1+\Delta_2-(\Delta-J+2n_0)=0\,.
\label{eq: Mellin truncation}
\ee
This is because the Mellin space encodes all multi-twist families by Gamma function factors. This truncation also appears for multi-twist operators.
 
In this paper, we reverse the OPE analysis described above. We first analyze the OPE in the coordinate space and write down the most general ansatz of Mellin amplitudes compatible with the OPE analysis, followed by solving the ansatz using the Casimir equation. We will provide more details in the next section.

\section{Algorithm for AdS amplitudes in Mellin space}
\label{sec:Algorithm}
\subsection{The difference equation and the algorithm}

To compute a Witten diagram in Mellin space, we can formulate the Casimir equation using Mellin space. For any differential operator $\mathcal{D}$, we can define a difference operator to describe its action in the Mellin space
\be
 \mathcal{D} \langle\mathcal{O}_1\cdots\mathcal{O}_n\rangle =  \fft{\mathcal{N}}{(2\pi i)^{\fft{n(n-3)}{2}}}\int [d\delta_{ij}]\prod_{i<j}\fft{\Gamma(\delta_{ij})}{(-2X_i\cdot X_j)^{\delta_{ij}}}\hat{\mathcal{D}}\circ M(\delta_{ij})\,.
\ee
In this way, we write the equation \eqref{eq: cutting} as
\be
\hat{\mathcal{D}}_{12\cdots n}^{\Delta,J}\circ M_{V,P}^{1\cdots n\cdots}(\delta_{ij}) = M_{V-1,P-1}^{1\cdots n\cdots}(\delta_{ij})\,.\label{eq: cutting Mellin}
\ee
This equation is then a difference equation because as we act the differential operators on the Mellin representation of Witten diagrams, the action effectively shifts the Mellin variables. Below is the example for identical four-point case with external dimension $\Delta_\phi$ 
\be
& \hat{\mathcal{C}}_{12}\circ M(s,t)= \frac{1}{2} M(s,t) \left(-2 d s-8 \Delta _{\phi }^2+s^2+4 \Delta _{\phi } (s+2 t)-2 s t-2 t^2\right)\nn\\
&-2 \delta _{12}^2 M(s-2,t)-2 \delta _{12}^2 M(s-2,t+2)+2 \delta _{13}^2 M(s,t+2)+2 \delta _{14}^2 M(s,t-2)\,,\label{eq: C12 on 4pt}
\ee 
where for four-point we use simpler variables $(s,t)$
\be
\delta_{12}=\Delta_\phi-\fft{s}{2}\,,\quad \delta_{13}=\Delta_\phi-\fft{u}{2}\,,\quad \delta_{14}=\Delta_\phi-\fft{t}{2}\,.
\ee

Recursively applying \eqref{eq: cutting Mellin} to remove all propagators as we explained previously, we end up with a difference equation where the RHS of equation is pure polynomials in $\delta_{ij}$. In general, it is impossible to solve such difference equation. Fortunately, for tree-level Mellin amplitudes, we know the generic structure so that we can make most general ansatz with correct pole structures and power in $\delta_{ij}$ for yet-to-be-solved Mellin amplitudes by OPE analysis.  The difference action must remove all poles in the ansatz, which is constraining enough to solve the Mellin amplitudes in terms of recursion equations up to finite number of undetermined constants. This solution is fixed pure kinematically, thus capturing the Mellin transform of conformal blocks, i.e., Mack polynomials. These constants encode the dynamics and can be determined by the action of boundary Feynman rules in the Mellin space, i.e., the LHS of the difference equation. We summarize this procedure by following algorithm, in a bit abstract way, however, we will provide examples of four-point function in a moment\footnote{See \cite{Zhou:2018sfz,Bissi:2022mrs,Ma:2022ihn} for a similar algorithm in the position space by using the integrated vertex identities, which, however, requires that at least one vertex is three-point. Our algorithm below may also be useful for capturing the AdS bound states by modifying the ansatz for those tree-level Witten diagrams with more fancy analytic structures \cite{Ma:2022ihn}. We are grateful to Xinan Zhou for pointing this out to us.}.

\vspace{10pt}

{\bf Algorithm}
\begin{itemize}
\item[1.] Given a Witten diagram, write down the most general ansatz of Mellin amplitudes by clearly separating rational part and singular part
\be
M(\delta_{ij}) = \fft{\text{rational polynomials}}{\text{poles}} + \text{rational polynomials}\,.
\ee
The part with poles should be irreducible so that it does not give rise to rational terms overlapping the pure polynomial part.  The maximal power of polynomials can be determined by performing dimensional analysis for the number of derivatives in the Witten diagram.

\item[2.] Write down the difference equation for the Witten diagram given the ansatz made in step $1$. Organize the difference equation so that the LHS is the action by (multiple) Casimir equation operators, while the RHS is the resulting contact terms given by the action of boundary Feynman rules.

\item[3.]  Require the LHS of difference equation poses no pole and solve the resulting recursion equations.

\item[4.] Substitute the solution of recursion equations into the difference equation to fully solve the Mellin amplitude of given Witten diagram.

\end{itemize}

\subsection{Examples of four-point}

In this section, we apply our algorithm to reproduce well-known four-point amplitudes with scalar, Yang-Mills and graviton exchanges \cite{Penedones:2010ue,Paulos:2011ie,Costa:2014kfa}, as a warm-up exercise and supplementary demo for the algorithm. 

We start with general single-trace exchange in four-point functions, see Fig \ref{fig: 4pt Witten}. For three-point vertices, any contracted derivatives can be removed by integration by parts and equations of motion, thus for spin-$J$ exchange, the vertex carries $J$ derivatives. Thus we write down the most general ansatz, as well-known in the literature
\be
M_{4}^{\Delta,J} =\sum_{m=0,1\cdots} \fft{Q_{Jm}(t)}{s-(\Delta-J)-2m} + R_{J-1}(s,t)\,,\label{eq: ansatz 4-pt}
\ee
where the maximal power of $Q_{Jm}$ and $R_{J-1}(s,t)$ is $J$ and $J-1$ respectively, namely
\be
Q_{Jm}(t)=\sum_{k=0}^J f_{Jkm}t^k\,,\quad R_{J-1}(s,t)=\sum_{a+b\leq J-1}R_{ab}s^a t^b\,.\label{eq: Q poly}
\ee
Applying \eqref{eq: C12 on 4pt}, we find the LHS of the difference equation is
\be
 \text{LHS:}\,\quad \hat{\mathcal{D}}_{12}^{\Delta,J}\circ M_{4}^{\Delta,J}=\sum_{m=0,1\cdots} \fft{\tilde{Q}_{J,m}(t)}{s-(\Delta-J)-2m} + \hat{\mathcal{D}}_{12}^{\Delta,J}\circ R_{J-1}(s,t)\,,
\ee
where
\begin{align}
& \tilde{Q}_{J,m}(t)=\left(s-\Delta _1-\Delta _2\right) \left(s-\Delta _3-\Delta _4\right) \left(Q_{J,m-1}(t)+Q_{J,m-1}(t+2)\right)\nn\\
& -\left(s+t-\Delta _1-\Delta _3\right) \left(s+t-\Delta _2-\Delta _4\right) Q_{J,m}(t+2)-\left(t-\Delta _2-\Delta _3\right) \left(t-\Delta _1-\Delta _4\right) Q_{J,m}(t-2)\nn\\
&+\Big(2 \left(\Delta  (\Delta -d)+J (d+J-2)+\Delta _3 \Delta _4\right)+s \left(2 d-\Delta _{\Sigma }+2 t\right)+\Delta _2 \left(\Delta _3+\Delta _4\right)\nn\\
&+\Delta _1 \left(2 \Delta _2+\Delta _3+\Delta _4\right)-s^2+2 t^2-2 t \Delta _{\Sigma }\Big)Q_{J,m}(t) \,.
\end{align}
This is equal to an effective contact diagram in the RHS, where the details are encoded by the specific bulk theories we consider. For RHS, we will provide the boundary Feynman rules for scalar, gluon and graviton exchanges case by case shortly.

Requiring the LHS is free of poles yields
\be
\tilde{Q}_{Jm}(t) = \big(y_0(t)+y_1(t)s\big)\big(s-(\Delta-J)-2m\big)\,.
\ee
Eliminating $y_0, y_1$ we then find the following recursion equation
\be
&-\left(\left(-\Delta _1-\Delta _2+2 m+\tau \right) \left(-\Delta _3-\Delta _4+2 m+\tau \right) \left(Q_{J,m-1}(t)+Q_{J,m-1}(t+2)\right)\right)\nn\\
&+\left(-\Delta _1-\Delta _3+2 m+t+\tau \right) \left(-\Delta _2-\Delta _4+2 m+t+\tau \right) Q_{J,m}(t+2)\nn\\
& +\Big(4 m (m-d)-2 \Delta _3 \Delta _4-\Delta _2 \left(\Delta _3+\Delta _4\right)-\Delta _1 \left(2 \Delta _2+\Delta _3+\Delta _4\right)+4J(1-J)-4 J \tau \nn\\
&+\Delta _{\Sigma } (2 m+\tau )+2 t \left(\Delta _1+\Delta _2+\Delta _3+\Delta _4-2 m-\tau \right)+4 m \tau -2 t^2-\tau ^2\Big) Q_{J,m}(t)\equiv 0\,.\label{eq: eq for mack}
\ee
This equation looks heavy, but it is precisely the equation for Mack polynomials provided by \cite{Costa:2012cb}\footnote{To identify with eq.~(131) in \cite{Costa:2012cb}, we simply swap $3\leftrightarrow 4$ and factorize out a normalization factor $Q_{J,m}^{\rm here}= \mathcal{P}_{Jm}Q_{J,m}^{\text{there}}$ where $\tau=\Delta-J$.}
We can easily solve Eq.~\eqref{eq: eq for mack} spin by spin, recalling \eqref{eq: Q poly}.
Notably, for a general $\Delta$, the solution reproduces the Mack polynomials \cite{Mack:2009mi,Costa:2012cb} up to normalization. This can be easily understood, as \eqref{eq: ansatz 4-pt} with the Mack polynomials inserted is proportional to the Mellin transform of the conformal block $G^{\Delta,J}$, which obeys the Casimir equation
\be
\mathcal{D}^{\Delta,J}_{12} G^{\Delta,J}\equiv 0\,. \label{eq: eq for conf block}
\ee
This equation can only be satisfied if the LHS is also free of any poles. The matching in this case can be easily achieved by requiring that the normalization has zeros on double-twist operators to cancel the relevant poles in the Gamma functions inside the Mellin integral. This is because the complex integral of completely smooth and regular functions is vanishing. It is worth noting that for those spinning operators saturating the unitary bounds, such as conserved current and stress-tensor, the solution to the recursion equation contains more undetermined constants. These constants can be fixed by the matching procedure, and unsurprisingly, the results also agree with the Mack polynomials. We will consider explicit examples for scalar, gluon, and graviton exchanges, and we will observe this fact explicitly by studying gluon and graviton exchanges

\begin{itemize}
\item {\bf Scalar exchange}
\end{itemize}

For scalar exchange, we consider the following Lagrangian
\be
\mathcal{L}=-\fft{1}{2}\sum_{i=1}^4\big((\nabla\phi_i)^2-m_i^2\phi_i^2\big)-\fft{1}{2}\big((\nabla\phi)^2-m^2\phi^2\big) -g( \phi_1\phi_2\phi + \phi_3\phi_4\phi)\,.
\ee
The boundary Feynman rules then give rise to the trivial RHS and the difference equation is
\be
\hat{\mathcal{D}}_{12}^{\Delta,0}\circ M_{4}^{\Delta,0} = -g^2\,.
\ee
The Mack polynomial is solved by
\eq{Q_{0m}(t)= c\, \mathcal{P}_{0m}\,,
}
where $c$ is the normalization constant and $\mathcal{P}_{0m}$ is
\be
\mathcal{P}_{Jm}=\frac{\left(\frac{1}{2} \left(\tau -\Delta _1-\Delta _2+4\right)\right)_{m-1} \left(\frac{1}{2} \left(\tau -\Delta _3-\Delta _4+4\right)\right)_{m-1}}{(2)_{m-1} \left(-\frac{d }{2}+\Delta+2\right)_{m-1}}\,.\label{eq: P poch}
\ee
We substitute this solution back to \eqref{eq: ansatz 4-pt} for $J=0$ and $R_{-1}\equiv 0$, after performing the infinite sum over $m$ it shall match with the RHS. This procedure gives
\eqs{c_0=g^2 \frac{(2+\Delta-\Delta_1-\Delta_2)(2+\Delta-\Delta_3-\Delta_4)\Gamma(\frac{\Delta-d+\Delta_1+\Delta_2}{2})\Gamma(\frac{\Delta-d+\Delta_3+\Delta_4}{2})}{16\Gamma(2+\Delta-\frac{d}{2})\Gamma(\frac{\Delta_\Sigma -d}{2})}\,,
}
which perfectly matches with the result from directly doing Feynman integral in the bulk \cite{Penedones:2010ue}.

\begin{itemize}
\item {\bf Gluon exchange}
\end{itemize}

Let's then take gluon exchange between four scalars as example. We consider the following Lagrangian

\be
\mathcal{L}=-\fft{1}{2}\sum_{i=1,3}(|D\phi_i|^2-m_i^2 |\phi_i|^2)-\fft{1}{4g_{\rm YM}^2}F^a_{\mu\nu}F^{\mu\nu a}\,,\label{eq: L of YM}
\ee
where $\Delta_2=\Delta_1, \Delta_4=\Delta_3$. The boundary Feynman rules for this interaction can be easily worked out \cite{Herderschee:2022ntr} 
\eqs{
\hat{\mathcal{D}}_{12}^{d-1,1} \circ M_{4}^{d-1,1}&= f_s\big(\hat{L}_2\cdot \hat{L}_3 - \hat{L}_2\cdot \hat{L}_4\big)\circ 1\\
&=2f_s \left(d-2 \left(\Delta _1+\Delta _3\right)\right) \left(\Delta _1+\Delta _3-\frac{s}{2}-t\right)
\,,\label{eq: gluon 4pt eq}}
where $f_s=f^{12a}f^{a34}g_{\rm YM}^2$ is the colour factor multiplied with the Yang-Mills coupling. In fact, due to the antisymmetry nature of YM interaction, we can simply replace $\hat{L}_i\cdot\hat{L}_j$ by $\delta_{ij}$ up to appropriate overall factor. This property is actually hidden by using $(s,t)$.

The single-trace operator in this case is conserved current on CFT side with $\Delta=d-1, J=1$, and we find solution of \eqref{eq: eq for mack} with more unfixed coefficients
\be
f_{11m}=c_1 \mathcal{P}_{1m}\,,\quad f_{10m}=\big(c_2\, m-\frac{1}{2} c_1 (m-1) \left(d-2 \left(\Delta _1+\Delta _3+1\right)\right)\big) \mathcal{P}_{1m}\,.
\ee
To determine the coefficients $c_i, R_0$, we substitute the solution back to \eqref{eq: gluon 4pt eq} and resum the series. In the end we find
\eqs{R_0= -f_s\,,\quad c_2=\fft{1}{2}c_1(d-2(\Delta_1+\Delta_3))\,,\quad c_1= f_s\frac{\left(d-2 \Delta _1\right) \left(d-2 \Delta _3\right) \Gamma \left(\Delta _1\right) \Gamma \left(\Delta _3\right)}{4 \Gamma \left(\frac{d}{2}+1\right) \Gamma \left(-\frac{d}{2}+\Delta _1+\Delta _3\right)}\,.}
It is important to note that by this matching we solved $c_2$ in terms of $c_1$, which precisely reproduces the Mack polynomials with $\Delta=d-1,J=1$. After summing over all poles, it perfectly matches with \cite{Paulos:2011ie}.

\begin{itemize}
\item {\bf Graviton exchange}
\end{itemize}

We consider scalar theories minimally coupled to AdS graviton
\be
\mathcal{L}=-\fft{1}{2}\sum_{i=1,3}\Big(\nabla_\mu \phi_i \nabla_\nu \phi_i h^{\mu\nu}-\fft{1}{2}g_{\mu\nu}h^{\mu\nu}\big((\nabla\phi)^2-m_i^2\phi_i^2\big)\Big)\,.
\ee
The corresponding boundary Feynman rules gives \def\hL{\hat{L}}\cite{Herderschee:2022ntr} 
\be
&\hat{\mathcal{D}}_{12}^{d,2}\circ M_4^{d,2}=-8\pi G\Big(\fft{1}{2}\big(\hat{L}_1\cdot \hL_4\, \hL_2\cdot \hL_3+\hat{L}_1\cdot \hL_3\, \hL_2\cdot \hL_4-\hL_1\cdot\hL_2\, \hL_3\cdot \hL_4\big)
\nn\\
&+\Delta_1\tilde{\Delta}_1 \hL_3\cdot \hL_4+\Delta_3\tilde{\Delta}_3 \hL_1\cdot \hL_4-\fft{2\Delta_1\tilde{\Delta}_1\Delta_3\tilde{\Delta}_3 (d+1)}{d-1}\Big)\circ 1\,,\label{eq: graviton 4pt eq}
\ee
where $\tilde{\Delta}=d-\Delta$ is the shadow dimension.

Since the exchanged operator is stress-tensor, similar to conserved current, solution of \eqref{eq: eq for mack}  contains one more undetermined coefficients
\be
&f_{22m}=c_1 \mathcal{P}_{2m}\,,\quad f_{21m}=\big(c_2\, m+ c_1 (1-m) \left(d-2 \left(\Delta _1+\Delta _3+1\right)\right)\big) \mathcal{P}_{2m}\,,\nn\\
& f_{20m}=\frac{1}{4} \Big(2 c_2 m \left(d-2 \left(\Delta _1+\Delta _3+1\right)+2 m\right)+c_1 \Big(4 \Delta _1 \left(d (2 m-1)+\Delta _3 (2-4 m)+2 (m-1)^2\right)\nn\\
&+4 \Delta _3 \Big(d (2 m-1)+\Delta _3 (1-2 m)\nn+2 (m-1)^2\big)-\frac{(d+2 m-2) ((d-2) d (2 m-1)+4 m)}{d-1}\nn\\
&+\Delta _1^2 (4-8 m)\Big)\Big)\mathcal{P}_{2m}\,.
\ee
Resuming the series and solving \eqref{eq: graviton 4pt eq} yields
\be
& R_{0,0}=8\pi G(2(\Delta_1^2+\Delta_3^2+\Delta_1\Delta_3)-d(\Delta_1+\Delta_3))\,,\quad R_{0,1}=8\pi G(d-2(\Delta_1+\Delta_3))\,,\quad R_{1,0}=0\,,\nn\\
& c_1=\frac{2 \pi  G \left(d-2 \Delta _1\right) \left(d-2 \Delta _3\right) \Gamma \left(\Delta _1+2\right) \Gamma \left(\Delta _3+2\right)}{\left(\Delta _1+1\right) \left(\Delta _3+1\right) \Gamma \left(\frac{d}{2}+2\right) \Gamma \left(-\frac{d}{2}+\Delta _1+\Delta _3\right)}\,,\quad c_2= (d-2(\Delta_1+\Delta_3))c_1\,.
\ee
$c_1$ and $c_2$ solved above is also consistent with Mack polynomials with $\Delta=d,J=2$. Our result agrees with bulk Feynman rules  \cite{Costa:2014kfa} by changing conventions $s\rightarrow t, t\rightarrow s+\Delta_1+\Delta_3$.

\section{Higher-point Mellin amplitudes}
\label{sec:Higher}
In this section, we apply our algorithm to study higher point Mellin amplitudes. We focus on six-point snowflake topology, which is expected to encode building blocks for all other topologies. Since the implementation of the algorithm for higher points is a bit more intricate than four points, we start with the scalar exchange as a warm-up, which shall match with scalar Mellin Feynman rules constructed in \cite{Nandan:2011wc}. Following the route of the scalar case, we then solve the most general snowflake six-point amplitude with three gluon exchanges. Besides, we also report simplified examples (i.e., with special conformal and spacetime dimensions) for other six-point amplitudes and two examples for eight-point amplitudes.

\subsection{Warm-up: six-point snowflake with scalar exchanges}

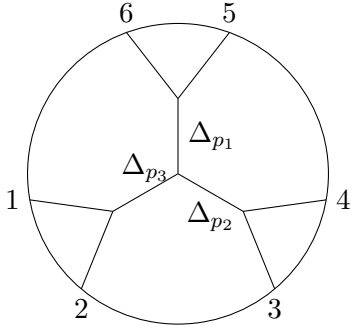
\begin{figure}[h]
\centering
 \begin{tikzpicture}
        \draw (0,0) circle (2 cm);
        
        \coordinate (1) at (-1.96962,-0.347296);
        \coordinate (2) at (-1.28558,-1.53209);
        \coordinate (A) at (-0.866025,-0.5);
        \coordinate (B) at (0.866025,-0.5);
        \coordinate (C) at (0,0);
        \coordinate (D) at (0,1);
         \coordinate (3) at (1.28558,-1.53209);
        \coordinate (4) at (1.96962,-0.347296);
          \coordinate (5) at (0.68404,1.87939);
        \coordinate (6) at (-0.68404,1.87939);
              
        \draw (1) -- (A);
        \draw (2) -- (A);
        \draw (A) -- (C) node[midway,above] {\(\Delta_{p_3}\)};
         \draw (3) -- (B);
        \draw (4) -- (B);
        \draw (B) -- (C) node[midway, below] {\(\Delta_{p_2}\)};
        \draw (5) -- (D);
        \draw (6) -- (D);
        \draw (D) -- (C) node[midway, right] {\(\Delta_{p_1}\)};
                  
        \fill (1)  node[left] {$1$};
        \fill (2) node[below] {$2$};
        \fill (3) node[below] {$3$};
         \fill (4) node[right] {$4$};
          \fill (5) node[above] {$5$};
         \fill (6) node[above] {$6$};
       \end{tikzpicture}
\caption{Six-point snowflake Witten diagram with scalar exchanges $\Delta_{p_1}$, $\Delta_{p_2}$ and $\Delta_{p_3}$.}
\label{fig: 6pt scalar snowflake}
\end{figure}

Consider snowflake scalar exchange Fig \ref{fig: 6pt scalar snowflake}, where we include the most general three-point vertex $\sum g_{ijk}\phi_i\phi_j\phi_k$. 
The corresponding equation is
\be
\hat{\mathcal{D}}_{56}^{\Delta_{p_1},0}\hat{\mathcal{D}}_{34}^{\Delta_{p_2},0}\hat{\mathcal{D}}_{12}^{\Delta_{p_3},0}\circ M_{12s,sss,s56,s34}=-g_{12,p_3}g_{34,p_2}g_{56,p_1}g_{p_1,p_2,p_3}\,.
\ee
Using the algorithm, we recursively solve this equation.

First, on the top level, we aim to solve the effective amplitude induced by
\be
M^{\rm eff}_{1234s,s56} := \hat{\mathcal{D}}_{34}^{\Delta_{p_2},0}\hat{\mathcal{D}}_{12}^{\Delta_{p_3},0}\circ M_{12s,sss,s56,s34}\,,\quad \hat{\mathcal{D}}_{56}^{\Delta_{p_1},0}\circ
M^{\rm eff}_{1234s,s56} =-\hat{g}\,
\ee
where $\hat{g}=g_{12,p_3}g_{34,p_2}g_{56,p_1}g_{p_1,p_2,p_3}$. See Fig \ref{fig: top level 6pt scalar} below.  
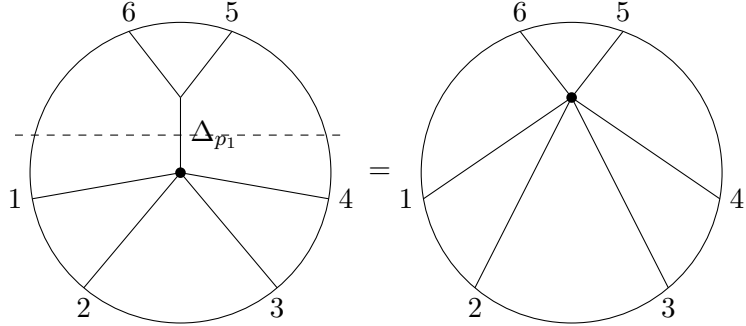
\begin{figure}[h]
\centering
 \begin{tikzpicture}
        \draw (0,0) circle (2 cm);
        
        \coordinate (1) at (-1.96962,-0.347296);
        \coordinate (2) at (-1.28558,-1.53209);
        \coordinate (A) at (-2.2,0.5);
        \coordinate (B) at (2.2,0.5);
        \coordinate (C) at (0,0);
        \coordinate (D) at (0,1);
         \coordinate (3) at (1.28558,-1.53209);
        \coordinate (4) at (1.96962,-0.347296);
          \coordinate (5) at (0.68404,1.87939);
        \coordinate (6) at (-0.68404,1.87939);
              
        \draw (1) -- (C);
        \draw (2) -- (C);
       \draw (3) -- (C);
        \draw (4) -- (C);
        \draw (5) -- (D);
        \draw (6) -- (D);
        \draw (D) -- (C) node[midway, right] {\(\Delta_{p_1}\)};
        \draw[dashed] (A) -- (B);
                  
        \fill (1)  node[left] {$1$};
        \fill (2) node[below] {$2$};
        \fill (3) node[below] {$3$};
         \fill (4) node[right] {$4$};
          \fill (5) node[above] {$5$};
         \fill (6) node[above] {$6$};
         \fill (C) circle (2pt);
         \node[anchor=west, right=2.3 cm of C] (formula) {\(\,=\)};  
        \draw (5.2,0) circle (2 cm);
        
        \coordinate (1p) at (3.23038,-0.347296);
        \coordinate (2p) at (3.91442,-1.53209);
        \coordinate (Dp) at (5.2,1);
         \coordinate (3p) at (6.48558,-1.53209);
        \coordinate (4p) at (7.16962,-0.347296);
          \coordinate (5p) at (5.88404,1.87939);
        \coordinate (6p) at (4.51596,1.87939);
              
        \draw (1p) -- (Dp);
        \draw (2p) -- (Dp);
       \draw (3p) -- (Dp);
        \draw (4p) -- (Dp);
        \draw (5p) -- (Dp);
        \draw (6p) -- (Dp);                  
        \fill (1p)  node[left] {$1$};
        \fill (2p) node[below] {$2$};
        \fill (3p) node[below] {$3$};
         \fill (4p) node[right] {$4$};
          \fill (5p) node[above] {$5$};
         \fill (6p) node[above] {$6$};
         \fill (Dp) circle (2pt);
       \end{tikzpicture}
\caption{The Casimir cut equation for the effective amplitudes $M^{\rm eff}_{1234s,s56}$, where the black dot refers to the effective vertices.}
\label{fig: top level 6pt scalar}
\end{figure}
This step makes no much difference from four-point case. The ansatz at this level is
\be
M^{\rm eff}_{1234s,s56} = \sum_{m}\fft{f^{\rm eff}_{56,m}}{\delta_{56}+m+\fft{\Delta_{p_1}-\Delta_5-\Delta_6}{2}}\,.
\ee
Similar to four-point example, we solve $f^{\rm eff}_{56,m}$ by simply requiring the cancelation of pole in $\delta_{56}$
\be
f^{\rm eff}_{56,m}= c^{\rm eff}_{56} \mathcal{P}_{0\Delta_{p_1}m}^{\sum_{i=1}^4\Delta_i,\Delta_5+\Delta_6,\Delta_{p_1}}\,,
\ee
where $\mathcal{P}_{J\tau m}^{a,b,c}$ is a slight modification of \eqref{eq: P poch}
\be
\mathcal{P}_{J\tau m}^{a,b,c} = \frac{\left(\frac{1}{2} \left(\tau -a+4\right)\right)_{m-1} \left(\frac{1}{2} \left(\tau -b+4\right)\right)_{m-1}}{(2)_{m-1} \left(-\frac{d }{2}+c+2\right)_{m-1}}\,.\label{eq: P poch mod}
\ee
This function does not depend on $J$ explicitly, the role of $J$-subscript is nothing but reminding us what spin is exchanged. We will see this factor is quite universal in our discussions, and we usually factorize it out to make the difference equation simpler to solve. Then by the top level matching, we obtain
\be
c^{\rm eff}_{56}=
\frac{\hat{g}\left(\Delta _{p_1}-\sum_{i=1}^4 \Delta_i+2\right) \left(\Delta _{p_1}-\Delta _5-\Delta _6+2\right) \Gamma \left(\frac{\sum_{i=1}^4\Delta_i+\Delta _{p_1}-d}{2} \right) \Gamma \left(\frac{\Delta _5+\Delta _6+\Delta _{p_1}-d}{2}\right)}{16\, \Gamma \left(\frac{\Delta_\Sigma-d}{2} \right) \Gamma \left(-\frac{d}{2}+\Delta _{p_1}+2\right)}\,.\label{eq: scalar 1234-to-56}
\ee
We can easily check this result matches with scalar Mellin Feynman rules. (we review the scalar Mellin Feynman rules in appendix \ref{sec: rules2}).

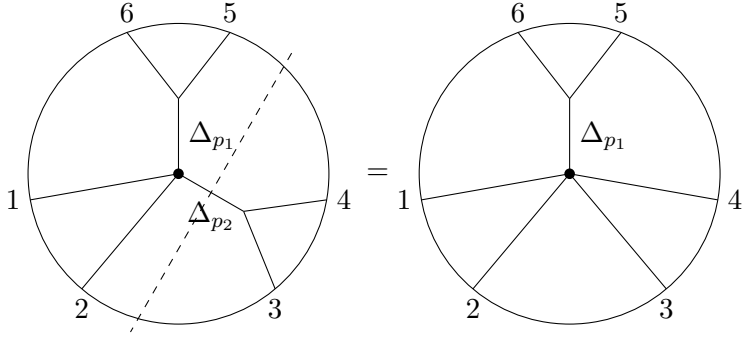
\begin{figure}[h]
\centering
 \begin{tikzpicture}
        \draw (0,0) circle (2 cm);
        
        \coordinate (1) at (-1.96962,-0.347296);
        \coordinate (2) at (-1.28558,-1.53209);
        \coordinate (A1) at (1.50423,1.6054);
         \coordinate (A2) at (-0.638202,-2.1054);
        \coordinate (B) at (0.866025,-0.5);
        \coordinate (C) at (0,0);
        \coordinate (D) at (0,1);
         \coordinate (3) at (1.28558,-1.53209);
        \coordinate (4) at (1.96962,-0.347296);
          \coordinate (5) at (0.68404,1.87939);
        \coordinate (6) at (-0.68404,1.87939);
              
        \draw (1) -- (C);
        \draw (2) -- (C);
         \draw (3) -- (B);
        \draw (4) -- (B);
        \draw (B) -- (C) node[midway, below] {\(\Delta_{p_2}\)};
        \draw (5) -- (D);
        \draw (6) -- (D);
        \draw (D) -- (C) node[midway, right] {\(\Delta_{p_1}\)};
        \draw[dashed] (A1) -- (A2);
                  
        \fill (1)  node[left] {$1$};
        \fill (2) node[below] {$2$};
        \fill (3) node[below] {$3$};
         \fill (4) node[right] {$4$};
          \fill (5) node[above] {$5$};
         \fill (6) node[above] {$6$};
         \fill (C) circle (2pt);
         \node[anchor=west, right=2.3 cm of C] (formula) {\(\,=\)};  
        \draw (5.2,0) circle (2 cm);
        
        \coordinate (1p) at (3.23038,-0.347296);
        \coordinate (2p) at (3.91442,-1.53209);
         \coordinate (Cp) at (5.2,0);
        \coordinate (Dp) at (5.2,1);
         \coordinate (3p) at (6.48558,-1.53209);
        \coordinate (4p) at (7.16962,-0.347296);
          \coordinate (5p) at (5.88404,1.87939);
        \coordinate (6p) at (4.51596,1.87939);
              
        \draw (1p) -- (Cp);
        \draw (2p) -- (Cp);
       \draw (3p) -- (Cp);
        \draw (4p) -- (Cp);
        \draw (5p) -- (Dp);
        \draw (6p) -- (Dp);   
       \draw (Cp) -- (Dp) node[midway, right] {\(\Delta_{p_1}\)};               
        \fill (1p)  node[left] {$1$};
        \fill (2p) node[below] {$2$};
        \fill (3p) node[below] {$3$};
         \fill (4p) node[right] {$4$};
          \fill (5p) node[above] {$5$};
         \fill (6p) node[above] {$6$};
         \fill (Cp) circle (2pt);
       \end{tikzpicture}
\caption{The Casimir cut equation for the effective amplitudes $M^{\rm eff}_{12ss,s34,s56}$.}
\label{fig: 6pt sec scalar}
\end{figure}

The next step is then to solve the equation for an effective amplitude $M^{\rm eff}_{34s,56s,12ss}$ as shown in Fig \ref{fig: 6pt sec scalar}
\be
M^{\rm eff}_{12ss,s34,s56} :=\hat{\mathcal{D}}_{12}^{\Delta_{p_3},0}\circ M_{12s,sss,s56,s34}\,,\quad \hat{\mathcal{D}}_{34}^{\Delta_{p_2},0}\circ M^{\rm eff}_{12ss,s34,s56}=M^{\rm eff}_{1234s,s56} \,,
\ee
where the ansatz is
\be
M^{\rm eff}_{12ss,s34,s56} = \sum_{n,m}\fft{f^{\rm eff}_{34,n,m}}{\big(\delta_{34}+n+\fft{\Delta_{p_2}-\Delta_3-\Delta_4}{2}\big)\big(\delta_{56}+m+\fft{\Delta_{p_1}-\Delta_5-\Delta_6}{2}\big)}\,.
\ee
Now it is more subtle to require the disappearance of poles in $\delta_{34}$ after the action of $\mathcal{D}_{34}^{\Delta_{p_2},0}$. The action of $\mathcal{D}_{34}^{\Delta_{p_2},0}$ leads to a reducible expression that can be further decomposed into irreducible partial fractions with respect to $\delta_{56}$, namely
\begin{align}
\hat{\mathcal{D}}_{34}^{\Delta_{p_2},0}\circ M^{\rm eff}_{12ss,s34,s56}=\sum_{nm}\Big(\ft{4 \left(f^{\text{eff}}_{34,n-1,m}-f^{\text{eff}}_{34,n-1,m-1}\right)\delta _{34}}{\delta_{34}+n+\fft{\Delta_{p_2}-\Delta_3-\Delta_4}{2}}-\ft{\mathcal{F}_{nm}(\delta_{34})}{\big(\delta_{34}+n+\fft{\Delta_{p_2}-\Delta_3-\Delta_4}{2}\big)\big(\delta_{56}+m+\fft{\Delta_{p_1}-\Delta_5-\Delta_6}{2}\big)}\Big)\,,\label{eq: D34 scalar}
\end{align}
where
\begin{align}
& \mathcal{F}_{nm}(\delta_{34})=(2 \delta _{34}-\Delta _3-\Delta _4+\Delta _{p_2}) (-d-2 \delta _{34}+\Delta _3+\Delta _4+\Delta _{p_2}) f^{\text{eff}}_{34,n,m}\nn\\
&+2 \delta _{34} ((2 \delta _{34}+\Delta _{13}+\Delta _{24}+2 m+\Delta _{p_1}) f^{\text{eff}}_{34,n-1,m}+(\Delta _5+\Delta _6-2 m-\Delta _{p_1}) f^{\text{eff}}_{34,n-1,m-1})\,.
\end{align}
Notably, the first term in \eqref{eq: D34 scalar} completely vanishes, since it allows to shift $m-1$ to $m$ without changing the result. Then kinematically $\mathcal{F}_{nm}(\delta_{34})$ should cancel pole in $\delta_{34}$. We factorize an appropriate factor as follows
\be
f^{\rm eff}_{34,nm} = \mathcal{P}_{0\Delta_{p_2}n}^{\Delta_{p_1}+\Delta_1+\Delta_2+2m,\Delta_3+\Delta_4,\Delta_{p_2}} \fft{\big(\fft{1}{2}(4+\Delta_{p_1}-\Delta_5-\Delta_6)\big)_{m-1}}{\big(\fft{1}{2}(\Delta_{p_1p_2}+\Delta_1+\Delta_2)\big)_{m-1}}\hat{f}^{\rm eff}_{34,nm}\,.
\ee
Then we find an extremely simple recursion equation for $\hat{f}$
\be
 \hat{f}^{\rm eff}_{34,nm} +\hat{f}^{\rm eff}_{34,n-1,m-1}- \hat{f}^{\rm eff}_{34,n-1,m}\equiv 0\,,
 \label{eq: eq for f34}
\ee
which can be simply solved by
\be
\hat{f}^{\rm eff}_{34,nm}= \sum_{k=0}^m c_{34,k} \fft{(-n)_{m-k}}{(1)_{m-k}}\,.\label{eq: sol of f34}
\ee
We note $\hat{f}^{\rm eff}_{34,nm}$ is a universal building block that also appears in other diagrams such as gluon exchange. The dynamics now are encoded in the undetermined coefficients $c_{34,k}$, requiring efforts to do the matching procedure. Similar to the four-point case, in the matching procedure we have to sum infinite series for $n$, as a result, we obtain a lot of hypergeometric functions which can be, although not manifestly, easily reduced to simple Gamma functions.  We emphasize again what we aim to match is \eqref{eq: scalar 1234-to-56}, which is itself an infinite series. The simplest way we can imagine is to do matching $m$ by $m$, thanks to the denominator $\delta_{56}+m+\cdots$. Not surprisingly ultimately we find a unique solution of $c_{34,k}$.
\begin{align}
& c_{34,k}= -c^{\rm eff}_{56} \,(k+1)\mathcal{P}_{0,\Delta_{p_1}-1,k+1}^{\Delta_{p_2}+1+\Delta_1+\Delta_2,\Delta_{p_2}+3-\Delta_1
-\Delta_2,\Delta_{p_1}-1}\times \nn\\
& \ft{\left(-\Delta _3-\Delta _4+\Delta _{p_2}\right) \left(-\Delta _3-\Delta _4+\Delta _{p_2}+2\right) \left(-d+2 \Delta _{p_1}+2\right) \Gamma \left(\frac{1}{2} \left(-d+\Delta _3+\Delta _4+\Delta _{p_2}+2\right)\right) \Gamma \left(\frac{1}{2} \left(-d+\Delta _1+\Delta _2+\Delta _{p_1}+\Delta _{p_2}\right)\right)}{2 \left(-\Delta _1-\Delta _2-\Delta _3-\Delta _4+\Delta _{p_1}+2\right) \Gamma \left(\frac{1}{2} \left(-d+\Delta _1+\Delta _2+\Delta _3+\Delta _4+\Delta _{p_1}\right)\right) \Gamma \left(-\frac{d}{2}+\Delta _{p_2}+2\right)}\,.
\end{align}
It is easy to check with \eqref{eq: scalar 1234-to-56} that the result at this level is symmetric in $(m,\Delta_{5,6})\leftrightarrow (n,\Delta_{3,4})$, and also perfectly matches with the scalar Mellin Feynman rules.

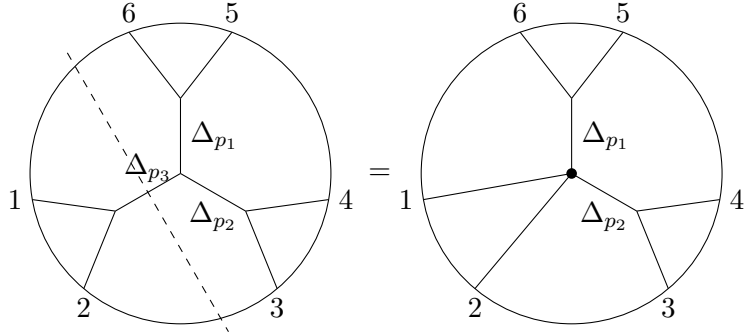
\begin{figure}[h]
\centering
 \begin{tikzpicture}
        \draw (0,0) circle (2 cm);
        
        \coordinate (1) at (-1.96962,-0.347296);
        \coordinate (2) at (-1.28558,-1.53209);
        \coordinate (A) at (-0.866025,-0.5);
        \coordinate (B) at (0.866025,-0.5);
        \coordinate (C) at (0,0);
        \coordinate (D) at (0,1);
         \coordinate (3) at (1.28558,-1.53209);
        \coordinate (4) at (1.96962,-0.347296);
          \coordinate (5) at (0.68404,1.87939);
        \coordinate (6) at (-0.68404,1.87939);
         \coordinate (A1) at (-1.50423,1.6054);
         \coordinate (A2) at (0.638202,-2.1054);
              
        \draw (1) -- (A);
        \draw (2) -- (A);
        \draw (A) -- (C) node[midway,above] {\(\Delta_{p_3}\)};
         \draw (3) -- (B);
        \draw (4) -- (B);
        \draw (B) -- (C) node[midway, below] {\(\Delta_{p_2}\)};
        \draw (5) -- (D);
        \draw (6) -- (D);
        \draw (D) -- (C) node[midway, right] {\(\Delta_{p_1}\)};
        \draw[dashed] (A1) -- (A2);
                  
        \fill (1)  node[left] {$1$};
        \fill (2) node[below] {$2$};
        \fill (3) node[below] {$3$};
         \fill (4) node[right] {$4$};
          \fill (5) node[above] {$5$};
         \fill (6) node[above] {$6$};
          \node[anchor=west, right=2.3 cm of C] (formula) {\(\,=\)};  
        \draw (5.2,0) circle (2 cm);
        
        \coordinate (1p) at (3.23038,-0.347296);
        \coordinate (2p) at (3.91442,-1.53209);
         \coordinate (Bp) at (6.066025,-0.5);
         \coordinate (Cp) at (5.2,0);
        \coordinate (Dp) at (5.2,1);
         \coordinate (3p) at (6.48558,-1.53209);
        \coordinate (4p) at (7.16962,-0.347296);
          \coordinate (5p) at (5.88404,1.87939);
        \coordinate (6p) at (4.51596,1.87939);
              
        \draw (1p) -- (Cp);
        \draw (2p) -- (Cp);
       \draw (3p) -- (Bp);
        \draw (4p) -- (Bp);
        \draw (5p) -- (Dp);
        \draw (6p) -- (Dp);   
       \draw (Cp) -- (Dp) node[midway, right] {\(\Delta_{p_1}\)}; 
        \draw (Cp) -- (Bp) node[midway, below] {\(\Delta_{p_2}\)};               
        \fill (1p)  node[left] {$1$};
        \fill (2p) node[below] {$2$};
        \fill (3p) node[below] {$3$};
         \fill (4p) node[right] {$4$};
          \fill (5p) node[above] {$5$};
         \fill (6p) node[above] {$6$};
         \fill (Cp) circle (2pt);
       \end{tikzpicture}
\caption{The Casimir cut equation for $M_{12s,sss,s56,s34}$ that defines the effective amplitude $M^{\rm eff}_{12ss,s34,s56}$.}
\label{fig: 6pt scalar finnal}
\end{figure}

Finally, at the bottom level, we have equation
\be
\hat{\mathcal{D}}_{12}^{\Delta_{p_3},0}\circ M_{12s,sss,s56,s34} = M^{\rm eff}_{12ss,s34,s56}\,,
\ee
as shown in Fig \ref{fig: 6pt scalar finnal}. 
The ansatz as usual is given by
\be
M_{12s,sss,s56,s34} = \sum_{pnm}\fft{f_{12,pnm}}{\big(\delta_{12}+p+\fft{\Delta_{p_3}-\Delta_1-\Delta_2}{2}\big)\big(\delta_{34}+n+\fft{\Delta_{p_2}-\Delta_3-\Delta_4}{2}\big)\big(\delta_{56}+m+\fft{\Delta_{p_1}-\Delta_5-\Delta_6}{2}\big)}\,.
\ee
There is no more subtlety at this level. Similarly, we find it is instructive to factorize $f_{12,pnm}$ to
\be
f_{12,pnm} = \mathcal{P}_{0\Delta_{p_3}p}^{\Delta_{p_1}+\Delta_{p_2}+2(m+n),\Delta_1+\Delta_2,\Delta_{p_3}}\ft{\left(\frac{1}{2} \left(-\Delta _5-\Delta _6+\Delta _{p_1}+4\right)\right)_{m-1} \left(\frac{1}{2} \left(-\Delta _3-\Delta _4+\Delta _{p_2}+4\right)\right)_{n-1}}{\left(\frac{1}{2} \left(\Delta _{p_1}+\Delta _{p_2}-\Delta _{p_3}\right)\right)_{m+n-1}}\hat{f}_{12,pnm}\,.
\ee
The cancelation of $\delta_{12}$ pole then provides another universal equation
\be
\hat{f}_{12,pnm}+\hat{f}_{12,p-1,n,m-1}+\hat{f}_{12,p-1,n-1,m}-\hat{f}_{12,p-1,n,m}\equiv 0\,.
\ee
The solution to this recursion equation is
\be
\hat{f}_{12,pnm} = \sum_{k_1=0}^n \sum_{k_2=0}^m c_{12,k_1k_2}\fft{(-p)_{n+m-k_1-k_2}}{(1)_{n-k_1}(1)_{m-k_2}}\,.\label{eq: sol of f12}
\ee
By matching to $M^{\rm eff}_{12ss,s34,s56}$, we find
\begin{align}
& c_{12,k_1,k_2}=c_{34,0}\sum_{q}(-1)^{k_1+k_2+1} \left(-\Delta _1-\Delta _2+\Delta _{p_3}+2\right) \left(-d+2 \Delta _{p_2}+2\right) \left(-k_1\right)_q \times\nn\\
& \ft{\left(\frac{1}{2} \left(-d+\Delta _1+\Delta _2+\Delta _{p_1}+\Delta _{p_2}\right)\right)_{k_1} \left(\frac{1}{2} \left(-d+2 k_1+\Delta _1+\Delta _2+\Delta _{p_1}+\Delta _{p_2}\right)\right)_{k_2} \left(\frac{1}{2} \left(-d+\Delta _{p_1}+\Delta _{p_2}+\Delta _{p_3}\right)\right)_q \left(\frac{1}{2} \left(-2 q+\Delta _{p_1p_3}-\Delta _{p_3}+2\right)\right)_{k_2}}{16 k_1! k_2! q! \left(\frac{1}{2} \left(-d+2 \Delta _{p_1}+2\right)\right)_{k_2} \left(\frac{1}{2} \left(-d+2 \Delta _{p_2}+2\right)\right)_q}\,,
\end{align}
which then gives us the complete answer of the snowflake six-point Mellin amplitudes with scalar exchange, as expected by scalar Mellin Feynman rules.

\subsection{Six-point snowflake with gluon exchanges}

\begin{figure}[h]
\centering
 \begin{tikzpicture}
        \draw (0,0) circle (2 cm);
        
        \coordinate (1) at (-1.96962,-0.347296);
        \coordinate (2) at (-1.28558,-1.53209);
        \coordinate (A) at (-0.866025,-0.5);
        \coordinate (B) at (0.866025,-0.5);
        \coordinate (C) at (0,0);
        \coordinate (D) at (0,1);
         \coordinate (3) at (1.28558,-1.53209);
        \coordinate (4) at (1.96962,-0.347296);
          \coordinate (5) at (0.68404,1.87939);
        \coordinate (6) at (-0.68404,1.87939);
              
        \draw (1)-- (A);
        \draw (2) -- (A);
        \draw[decorate, decoration={coil, aspect=0, segment length=5pt, amplitude=4pt}]  (A) -- (C) ;
          \draw  (3) -- (B);
        \draw (4) -- (B);
         \draw[decorate, decoration={coil, aspect=0, segment length=5pt, amplitude=4pt}]  (B) -- (C); 
         \draw (5) -- (D);
        \draw (6) -- (D);
        \draw[decorate, decoration={coil, aspect=0, segment length=5pt, amplitude=4pt}]  (D) -- (C);
                  
        \fill (1)  node[left] {$1$};
        \fill (2) node[below] {$2$};
        \fill (3) node[below] {$3$};
         \fill (4) node[right] {$4$};
          \fill (5) node[above] {$5$};
         \fill (6) node[above] {$6$};
       \end{tikzpicture}
\caption{Six-point snowflake Witten diagram with gluon exchanges. The coil lines represent gluon, giving rise to single-trace conserved current with $(\Delta=d-1,J=1)$.}
\label{fig: 6pt gluon snowflake}
\end{figure}
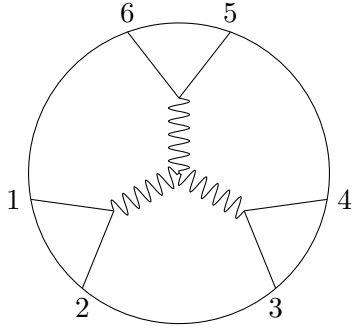

Educated by the scalar snowflake exercise, we are ready to report the snowflake gluon result as shown in Fig \ref{fig: 6pt gluon snowflake}.
This diagram can be generated by the Lagrangian in \eqref{eq: L of YM} appended by additional scalar $\phi_5$. As constructed in \cite{Herderschee:2022ntr} (with a slight generalization to any mass), the boundary Feynman rule for this diagram is (under the action $\hat{\mathcal{D}}_{56}^{d-1,1}\hat{\mathcal{D}}_{34}^{d-1,1}\hat{\mathcal{D}}_{12}^{d-1,1}\circ M_{12g,ggg,g56,g34}$ \begin{align}\def\hL{\hat{L}}
& -\fft{C_{\rm snow}}{2}\Big(\big(\hL_{13}\hL_{26}-\hL_{13}\hL_{25}+\hL_{13}\hL_{45}-\hL_{13}\hL_{46}+\hL_{23}\hL_{46}-\hL_{23}\hL_{45}\big) - \big(3\leftrightarrow 4\big) \nn\\
&- \big(3\leftrightarrow 5, 4\leftrightarrow 6\big) - \big(3\leftrightarrow 6, 4\leftrightarrow 5\big) \Big)\circ 1\,,
\end{align}
where $C_{\rm snow}$ encodes the color factor and Yang-Mills coupling
\be
C_{\rm snow}= f^{12g_1}f^{34g_2}f^{56g_3}f^{g_1g_2g_3}g_{\rm YM}^4\,.
\ee
This action, as evaluated in Mellin space, gives rise to
\begin{align}
&P_{\rm snow}=2C_{\rm snow}\tilde{\Delta}_\Sigma(2-\tilde{\Delta}_\Sigma)\Big(\big(\delta_{13}\delta_{26}-\delta_{13}\delta_{25}+\delta_{13}\delta_{45}-\delta_{13}\delta_{46}+\delta_{23}\delta_{46}-\delta_{23}\delta_{45}\big) - \big(3\leftrightarrow 4\big) \nn\\
&- \big(3\leftrightarrow 5, 4\leftrightarrow 6\big) - \big(3\leftrightarrow 6, 4\leftrightarrow 5\big)\Big) \,.
\end{align}
As we already observed in four-point case, this function simply replaces $\hL_i\cdot\hL_j$ by $\delta_{ij}$ in the boundary Feynman rule, thanks to the gauge group. Furthermore, we emphasize that this ``flat extension'' property enables us to find an extremely simple ansatz, putting gluon and scalar exchange on the same level
\begin{align}
&M^{\rm eff}_{1234g,g56}= \sum_m \fft{f_{56g,m}P_{\rm snow}}{\delta_{56}+m+\fft{d-2-2\Delta_5}{2}}\,,\nn\\
&M^{\rm eff}_{12gg,g34,g56} =  \sum_{nm}\fft{f_{34g,nm}P_{\rm snow}}{\big(\delta_{34}+n+\fft{d-2-2\Delta_3}{2}\big)\big(\delta_{56}+m+\fft{d-2-2\Delta_5}{2}\big)}\,,\nn\\
&M_{12g,ggg,g56,g34} = \sum_{pnm}\fft{f_{12g,pnm}P_{\rm snow}}{\big(\delta_{12}+p+\fft{d-2-2\Delta_1}{2}\big)\big(\delta_{34}+n+\fft{d-2-2\Delta_3}{2}\big)\big(\delta_{56}+m+\fft{d-2-2\Delta_5}{2}\big)}\,,
\end{align}
where $M^{\rm eff}$ is defined through Casimir cuts, similar to those in the previous section by replacing $s$ by $g$, $J=0$ by $J=1$, specifying $\Delta_{p_i}=d-1$. Generally, if we are thinking about arbitrary spin-$1$ exchange, we have to make ansatz containing the most general polynomials up to the degree of the number of involved derivatives, which include a large number of yet-to-be-solved functions. The nice property of YM allows us to reduce to only one function in the ansatz.

The top level makes no much difference from the scalar exchange, we arrive at
\be
f^{\rm eff}_{56g,m} = c_{56g}^{\rm eff}\mathcal{P}_{1,d-1,m}^{2(\Delta_1+\Delta_3+1),2\Delta_5,d-1}\,,
\ee
for which the matching now trivially determines
\be
c_{56g}^{\rm eff}=-\frac{\left(d-2 \Delta _1-2 \Delta _3-2\right) \left(d-2 \Delta _5\right) \Gamma \left(\Delta _1+\Delta _3+1\right) \Gamma \left(\Delta _5\right)}{16 \Gamma \left(\frac{d}{2}+1\right) \Gamma \left(-\frac{d}{2}+\Delta _1+\Delta _3+\Delta _5+2\right)}\,.
\ee
At the next level, we obtain
\be
f^{\rm eff}_{34g,nm} = \mathcal{P}_{1,d-2,n}^{d+2\Delta_1+2m,2\Delta_3,d-1} \fft{\big(1+\fft{d}{2}-\Delta_5)\big)_{m-1}}{\big(1+\Delta_1)\big)_{m-1}}\hat{f}^{\rm eff}_{34,nm}\,,
\ee
where $\hat{f}^{\rm eff}_{34,nm}$ satisfies eq.~\eqref{eq: eq for f34} and is solved by \eqref{eq: sol of f34} by replacing $c^{\rm eff}_{34,k}$ by $c^{\rm eff}_{34g,k}$
\bea
c^{\rm eff}_{34g,k}=c^{\rm eff}_{56g} (k+1)\mathcal{P}_{1,d-2,k+1}^{d+2+2\Delta_1,d-2\Delta_1,d-2}\times \frac{\left(d-2 \Delta _3\right) \Gamma \left(\Delta _3\right) \Gamma \left(\frac{d}{2}+\Delta _1\right)}{4 \left(\Delta _1+\Delta _3\right) \left(d-2 \left(\Delta _1+\Delta _3+1\right)\right) \Gamma \left(\frac{d}{2}\right) \Gamma \left(\Delta _1+\Delta _3\right)}\,.
\cr &&
\eea
Finally, we move to the bottom level and obtain the solution of the target amplitude
\be
f_{12,pnm} = \mathcal{P}_{0,d-2,p}^{2(d+m+n),2\Delta_1,d-1}\fft{\left(\frac{1}{2} \left(2+d-2\Delta_5\right)\right)_{m-1} \left(\frac{1}{2} \left(2+d-2\Delta_3\right)\right)_{n-1}}{\left(1+\fft{d}{2}\right)_{m+n-1}}\hat{f}_{12,pnm}\,,\label{eq: f12pnm snow}
\ee
where $\hat{f}_{12,pnm}$ is universally solved by \eqref{eq: sol of f12}, while the matching changes $c_{12,k_1k_2}$ to a much more compact expression $c_{12g,k_1k_2}$
\be
c_{12g,k_1k_2}=c^{\rm eff}_{56g} \ft{\left(d-2 \Delta _1\right) \left(d-2 \Delta _3\right) \Gamma (d-1) \Gamma \left(\Delta _1\right) \Gamma \left(\Delta _3\right) \left(\frac{2-d}{2}\right)_{k_2} \left(\frac{1}{2} \left(d+2 k_2\right)\right)_{k_1} \Gamma \left(\frac{d}{2}-k_2\right) \Gamma \left(-\frac{d}{2}+k_1-k_2+1\right)}{32 k_1! k_2! \left(d-2 \Delta _1-2 \Delta _3-2\right) \Gamma \left(\frac{d}{2}\right) \Gamma \left(\frac{d+2}{2}\right) \Gamma \left(\Delta _1+\Delta _3+1\right) \Gamma \left(-\frac{d}{2}-k_2+1\right) \Gamma \left(\frac{d}{2}+k_1-k_2\right)}\,.
\label{eq: snow coeff}
\ee

We find that our result can be rewritten in a format analogous to the scalar Mellin Feynman rules \cite{Nandan:2011wc}. We should slightly modify the vertex function (see \eqref{eq: vertex func}) to count the number of derivatives
\be
V_{m_1 \ldots m_k}^{N,\Delta_1 \ldots \Delta_k}=\sum_{n_1=0}^{m_1} \cdots \sum_{n_k=0}^{m_k} \Gamma\left(\frac{\sum_j\left(\Delta_j+2 n_j\right)-d+N}{2}\right) \prod_{j=1}^k \frac{\left(-m_j\right)_{n_j}}{n_{j} !\left(\Delta_j-\frac{d}{2}+1\right)_{n_j}}.
\ee
Then we verify that our result \eqref{eq: f12pnm snow} and \eqref{eq: snow coeff} can be rewritten by
\be
& f_{12,pnm}=-\frac{\left(\Delta _1-1\right) \left(\Delta _3-1\right) \left(\Delta _5-1\right)}{(d-2)^3 \Gamma \left(-\frac{d}{2}+\Delta _1+\Delta _3+\Delta _5+2\right)}\times \nn\\
& V_{00m}^{1,\Delta_5-1,\Delta_5,d-2}V_{00n}^{1,\Delta_3-1,\Delta_3,d-2}V_{00p}^{1,\Delta_1-1,\Delta_1,d-2}V_{nmp}^{1,d-1,d-1,d-1}S_m^{d-2}S_n^{d-2}S_p^{d-2}\,.
\ee

\subsection{Other truncated six-point amplitudes}
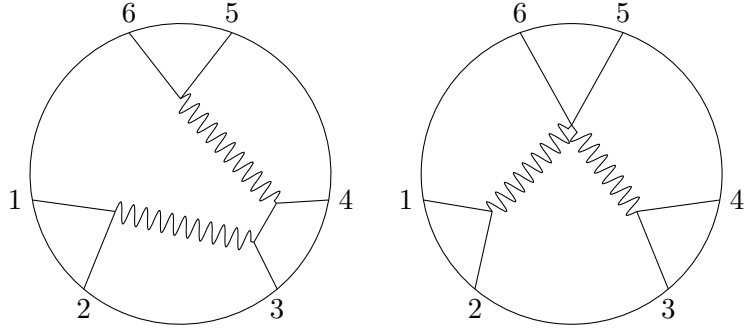
\begin{figure}[h]
\centering
 \begin{tikzpicture}
     \begin{scope}[rotate=240]
        \draw (0,0) circle (2 cm);
        
        \coordinate (1) at (-1.96962,-0.347296);
        \coordinate (2) at (-1.28558,-1.53209);
        \coordinate (A) at (-0.866025,-0.5);
        \coordinate (B) at (0.866025,-0.5);
        \coordinate (C) at (0,0);
        \coordinate (D) at (0,1);
         \coordinate (3) at (1.28558,-1.53209);
        \coordinate (4) at (1.96962,-0.347296);
          \coordinate (5) at (0.68404,1.87939);
        \coordinate (6) at (-0.68404,1.87939);
         \coordinate (D1) at (-0.3,1.3);  
         \coordinate (D2) at (0.3,1.3); 
         \coordinate (B1) at (0.866025,-0.5);   
        \draw (1)-- (A);
        \draw (2) -- (A);
        \draw[decorate, decoration={coil, aspect=0, segment length=5pt, amplitude=4pt}]  (A) -- (D1) ;
          \draw  (3) -- (B);
        \draw (4) -- (B);
         \draw[decorate, decoration={coil, aspect=0, segment length=5pt, amplitude=4pt}]  (B) -- (D2); 
         \draw (5) -- (D2);
        \draw (6) -- (D1);
         \draw (D2) -- (D1);
        
        \fill (1)  node[above] {$5$};
        \fill (2) node[above] {$6$};
        \fill (3) node[left] {$1$};
         \fill (4) node[below] {$2$};
          \fill (5) node[below] {$3$};
         \fill (6) node[right] {$4$};
   \end{scope}
        \draw (5.2,0) circle (2 cm);
        
        \coordinate (1p) at (3.23038,-0.347296);
        \coordinate (2p) at (3.91442,-1.53209);
         \coordinate (Bp) at (6.066025,-0.5);
         \coordinate (Cp) at (5.2,0);
        \coordinate (Dp) at (5.2,0.65);
         \coordinate (3p) at (6.48558,-1.53209);
        \coordinate (4p) at (7.16962,-0.347296);
          \coordinate (5p) at (5.88404,1.87939);
        \coordinate (6p) at (4.51596,1.87939);
            \coordinate (Ap) at (4.144,-0.5);
        \draw (1p) -- (Ap);
        \draw (2p) -- (Ap);
     \draw[decorate, decoration={coil, aspect=0, segment length=5pt, amplitude=4pt}]  (Ap) -- (Dp) ;
 \draw[decorate, decoration={coil, aspect=0, segment length=5pt, amplitude=4pt}]  (Bp) -- (Dp);
       \draw (3p) -- (Bp);
        \draw (4p) -- (Bp);
         \draw (5p) -- (Dp);
        \draw (6p) -- (Dp);      
        \fill (1p)  node[left] {$1$};
        \fill (2p) node[below] {$2$};
        \fill (3p) node[below] {$3$};
         \fill (4p) node[right] {$4$};
          \fill (5p) node[above] {$5$};
         \fill (6p) node[above] {$6$};
       \end{tikzpicture}
\caption{Six-point Witten diagram with gluon exchanges. The coil lines represent gluon, giving rise to single-trace conserved current with $(\Delta=d-1,J=1)$.}
\label{fig: 6pt gluons}
\end{figure}
For the Lagrangian \eqref{eq: L of YM} , we still have two diagrams for six-point amplitudes as shown in Fig \ref{fig: 6pt gluons}, see also \cite{Herderschee:2022ntr}. In this subsection, we apply our algorithm to efficiently calculate these two diagrams by focusing on even integer dimensions and presenting results for identical scalars with $\Delta=d/2+k$ for $k\in \mathbb{Z}$ so that the amplitudes are truncated. Our target is to demonstrate the efficiency of our algorithm for computing truncated amplitudes, thus we will not be too strict with the overall normalization that can be absorbed into the YM coupling. We find patterns analogous to scalar Mellin Feynman rules, which we believe the final expressions are also valid for other dimensions and identical scalars. Unfortunately, we are not able to conclude a general Mellin Feynman rules for Witten diagrams involving gluons from our results.

The boundary Feynman rules can be easily constructed. More generally, since we are not limited to massless scalar, we should slightly modify the boundary Feynman rules constructed in \cite{Herderschee:2022ntr}
\begin{align}
& \hat{\mathcal{D}}_{34}^{d-1,1}\hat{\mathcal{D}}_{12}^{d-1,1}\circ M_{12g,gg56,g34}=C_{gg}
\big(\hL_{13}-\hL_{23}+\hL_{24}-\hL_{14}\big)\circ 1\,.\nn\\
& \hat{\mathcal{D}}_{56}^{d-1,1}\hat{\mathcal{D}}_{123}^{\Delta,0}\hat{\mathcal{D}}_{12}^{d-1,1}\circ M_{12g,g3s,s4g,g56}=C_{ggs} \big(\hL_{11}-\hL_{22}+2\hL_{13}-2\hL_{23}\big)\big(\hL_{55}-\hL_{66}+2\hL_{45}-2\hL_{46}\big)\circ 1\,,
\end{align}
\footnote{Note that in general $\hL_{11}\neq \hL_{22}$, but when acting on the correlation functions with $\Delta_1=\Delta_2$ we have $\hL_{11}\circ 1=\hL_{22}\circ 1$, precisely embracing the spirit of the on-shell condition for amplitudes.}where the color factors are given by
\begin{align}
C_{gg}=g_{\rm YM}^4\, f^{12g_1}f^{34g_2}(f^{5b_1g_2}f^{6b_1g_1}+f^{5b_1g_1}f^{6b_1g_2})\,,\quad C_{ggs}=g_{\rm YM}^4 f^{12g_1}f^{23g_1}f^{24g_3}f^{56g_3}\,.
\end{align}
Following our algorithm, we write the ansatz in a format as follows
\begin{align}
& M_{12g,gg56,g34}=P_{gg} \sum_{n=0,m=0}\fft{f_{gg}^{mn}}{\left(\frac{d}{2}+\delta _{34}-\Delta +m-1\right) \left(\frac{d}{2}+\delta _{12}-\Delta +n-1\right)}\,,\nn\\
& M_{12g,g3s,s4g,g56}= P_{ggs} \sum_{m=0,n=0,p=0}\ft{f_{ggs}^{mnp}}{\left(\frac{d}{2}+\delta _{56}-\Delta +m-1\right) \left(\frac{d}{2}+\delta _{12}+\delta _{13}+\delta _{23}-\frac{3 \Delta }{2}+n-1\right) \left(\frac{d}{2}+\delta _{12}-\Delta +p-1\right)}\,,
\end{align}
where the kinematic factors are precisely given by the boundary Feynman rules
\begin{align}
& P_{gg}=2C_{gg}(6\Delta-d)(\delta_{13}-\delta_{14}-\delta_{23}+\delta_{24})\,,\nn\\
& P_{ggs}=16 C_{ggs} (6 \Delta -d) (6 \Delta-d +2)\times\nn\\
&\left(\delta _{13}-\delta _{23}\right) \left(-2 \delta _{12}-2 \delta _{13}-\delta _{14}-2 \delta _{15}-2 \delta _{23}-\delta _{24}-2 \delta _{25}-\delta _{34}-2 \delta _{35}+3 \Delta \right)\,.
\end{align}
We are then ready for solving $f_{gg}^{nm}$ and $f_{ggs}^{mnp}$. We compute these coefficients for $d=4,6,8,10$ and $\Delta=d/2+k$ for at most $k=5$. As expected, we find a general pattern similar to the scalar Mellin Feynman rules.
\begin{align}
& f_{gg}^{mn} = \ft{(\Delta-1)^2}{(d-2)^2\Gamma\big(3\Delta-\fft{d}{2}+1\big)}V_{00n}^{1;\Delta-1,\Delta,d-2}V_{00m}^{1;\Delta-1,\Delta,d-2}V_{00nm}^{0;\Delta,\Delta,d-1,d-1}S_m^{d-2}S_n^{d-2}\,,\nn\\
&f_{ggs}^{mnp}=-\ft{(\Delta-1)^2}{2(d-2)^2 \Gamma\big(3\Delta-\fft{d}{2}+2\big)} V_{00p}^{1;\Delta-1,\Delta,d-2}V_{00m}^{1;\Delta-1,\Delta,d-2}V_{0np}^{1;\Delta,\Delta,d-1}V_{0nm}^{1;\Delta,\Delta,d-1}S_{m}^{\Delta}S_{n}^{d-2}S_{p}^{d-2}\,,
\end{align}
It is straightforward to check these results for higher dimensions and higher $k$. We claim the results here should also be valid for general dimensions and identical scaling dimensions $\Delta$.

\subsection{Simple examples at eight-point}

In this subsection, we provide solutions for two examples of eight-point amplitudes involving gluon exchange. These examples are among the simplest diagrams at the eight-point level. In particular, we will consider an example that is not gauge invariant by itself, which requires a more cautious approach when dealing with the Casimir cut.

\subsubsection{Example I}

\begin{figure}[h]
\centering
 \begin{tikzpicture}
        \draw (0,0) circle (2 cm);
        
        \coordinate (2) at (-0.70,-1.87);
        \coordinate (1) at (-1.87,-0.70);
        \coordinate (A) at (-0.707,-0.707);
        \coordinate (B) at (0.707,-0.707);
        \coordinate (C) at (0,0);
       \coordinate (D) at (-0.707,0.707);
        \coordinate (E) at (0.707,0.707);
         \coordinate (3) at (0.70,-1.87);
        \coordinate (4) at (1.87,-0.707);
         \coordinate (5) at (1.87,0.70);
        \coordinate (6) at (0.70,1.87);
        \coordinate (7) at (-0.70,1.87);
        \coordinate (8) at (-1.87,0.70);
        \draw (1)-- (A);
        \draw (2) -- (A);
        \draw[decorate, decoration={coil, aspect=0, segment length=5pt, amplitude=4pt}]  (A) -- (C) ;
          \draw  (3) -- (B);
        \draw (4) -- (B);
         \draw[decorate, decoration={coil, aspect=0, segment length=5pt, amplitude=4pt}]  (B) -- (C); 
         \draw (5) -- (E);
        \draw (6) -- (E);
        \draw[decorate, decoration={coil, aspect=0, segment length=5pt, amplitude=4pt}]  (D) -- (C);
         \draw (7) -- (D);
        \draw (8) -- (D);
        \draw[decorate, decoration={coil, aspect=0, segment length=5pt, amplitude=4pt}]  (E) -- (C);
        
        \fill (1)  node[left] {$1$};
        \fill (2) node[below] {$2$};
        \fill (3) node[below] {$3$};
         \fill (4) node[right] {$4$};
          \fill (5) node[right] {$5$};
         \fill (6) node[above] {$6$};
         \fill (7) node[above] {$7$};
         \fill (8) node[left] {$8$};
       \end{tikzpicture}
\caption{The first example of eight-point Witten diagrams with gluon exchanges.}
\label{fig: 8pt gluon snowflake}
\end{figure}
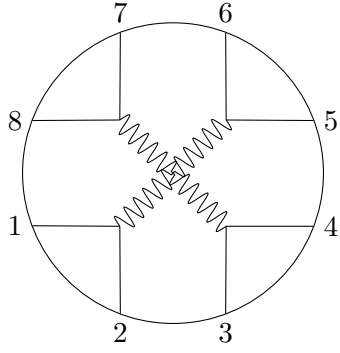

In Fig \ref{fig: 8pt gluon snowflake}, we consider an eight-point example, which is a comb diagram with only two derivatives.
The boundary Feynman rules for Fig \ref{fig: 8pt gluon snowflake} can be easily constructed as
\be
& \hat{\mathcal{D}}_{78}^{d-1,1}\hat{\mathcal{D}}_{56}^{d-1,1}\hat{\mathcal{D}}_{34}^{d-1,1}\hat{\mathcal{D}}_{12}^{d-1,1}\circ M_{12g,34g,56g,78g} \nn\\
&=C^a_{gggg}\left( \big(\hL_{15}-\hL_{16}+\hL_{26}-\hL_{25}\big)\big(\hL_{37}-\hL_{38}+\hL_{48}-\hL_{47}\big)-(5\leftrightarrow 7,6 \leftrightarrow 8) \right)\circ 1\,\nn\\
&+C^b_{gggg}\left( \big(\hL_{13}-\hL_{14}+\hL_{24}-\hL_{23}\big)\big(\hL_{57}-\hL_{58}+\hL_{68}-\hL_{67}\big)-(3\leftrightarrow 7,4 \leftrightarrow 8) \right)\circ 1\,\nn\\
&+C^c_{gggg}\left( \big(\hL_{13}-\hL_{14}+\hL_{24}-\hL_{23}\big)\big(\hL_{57}-\hL_{58}+\hL_{68}-\hL_{67}\big)-(3\leftrightarrow 5,4 \leftrightarrow 6) \right)\circ 1.
\label{eq: difference eq 8pt}
\ee
where $C^a_{gggg}$ encodes all the color factors and coupling constant and are given by
\eqs{
C^a_{gggg}&=g_{YM}^6 f^{12a} f^{34b} f^{56c} f^{78d} f^{abe} f^{cde}\,,\quad
C^b_{gggg}=g_{YM}^6 f^{12a} f^{34b} f^{56c} f^{78d} f^{ace} f^{bde}\,,\nn\\
C^c_{gggg}&=g_{YM}^6 f^{12a} f^{34b} f^{56c} f^{78d} f^{ade} f^{bce}\,.
}
With the flat extension properties stated before, we can write down the following simple ansatz
\begin{align}
 M_{12g,34g,56g,78g}=P_{gggg}\sum_{mnpq}\ft{f^{mnpq}_{gggg}}{(\frac{d}{2}+\delta_{12}-\Delta+m-1)(\frac{d}{2}+\delta_{34}-\Delta+n-1)(\frac{d}{2}+\delta_{56}-\Delta+p-1)(\frac{d}{2}+\delta_{78}-\Delta+q-1)}\,,
\end{align}
where
\eqs{
    P_{gggg}=&4(8\Delta-d)^2 \Big[C^a_{gggg}\left ((\delta_{15}-\delta_{16}+\delta_{26}-\delta_{25})(\delta_{37}-\delta_{38}+\delta_{48}-\delta_{47})-(5\leftrightarrow 7,6 \leftrightarrow 8) \right)\\
    &+C^b_{gggg}\left ((\delta_{13}-\delta_{14}+\delta_{24}-\delta_{23})(\delta_{57}-\delta_{58}+\delta_{68}-\delta_{67})-(3\leftrightarrow 7,4 \leftrightarrow 8) \right)\\
    &+C^c_{gggg}\left ((\delta_{13}-\delta_{14}+\delta_{24}-\delta_{23})(\delta_{57}-\delta_{58}+\delta_{68}-\delta_{67})-(3\leftrightarrow 5,4 \leftrightarrow 6) \right)\Big]\,.
}
We implemented the same strategy for solving the truncated case, and find the following general pattern for $f^{mnpq_{gggg}}$
\eqs{
f^{mnpq}_{gggg}=&\frac{(\Delta-1)^4}{(d-2)^4\Gamma(4\Delta-\frac{d}{2}+2)}V_{00m}^{1;\Delta-1,\Delta,d-2}V_{00n}^{1;\Delta-1,\Delta,d-2}V_{00p}^{1;\Delta-1,\Delta,d-2}V_{00q}^{1;\Delta-1,\Delta,d-2}\\
&\times V_{mnpq}^{0;d-1,d-1,d-1,d-1}S_m^{d-2}S_n^{d-2}S_p^{d-2}S_q^{d-2}\,.
}

\subsubsection{Example II: cautions with nonphysical modes}
\label{subsec: subtle non-gauge}

For the second example, we consider the eight-point amplitudes as shown in Fig \ref{fig: 8pt gluon caution} below.
\begin{figure}[h]
\centering
 \begin{tikzpicture}
        \draw (0,0) circle (2 cm);
        
        \coordinate (2) at (-0.70,-1.87);
        \coordinate (1) at (-1.87,-0.70);
        \coordinate (A) at (-0.707,-0.707);
        \coordinate (B) at (0.707,-0.707);
        \coordinate (C) at (0,0);
       \coordinate (D) at (-0.707,0.707);
        \coordinate (E) at (0.707,0.707);
         \coordinate (3) at (0.70,-1.87);
        \coordinate (4) at (1.87,-0.707);
         \coordinate (5) at (1.87,0.70);
        \coordinate (6) at (0.70,1.87);
        \coordinate (7) at (-0.70,1.87);
        \coordinate (8) at (-1.87,0.70);
        \draw (1)-- (A);
        \draw (2) -- (A);
        \draw[decorate, decoration={coil, aspect=0, segment length=5pt, amplitude=4pt}]  (A) -- (D) ;
          \draw  (3) -- (B);
        \draw (4) -- (B);
         \draw[decorate, decoration={coil, aspect=0, segment length=5pt, amplitude=4pt}]  (B) -- (E); 
         \draw (5) -- (E);
        \draw (6) -- (E);
        \draw[decorate, decoration={coil, aspect=0, segment length=5pt, amplitude=4pt}]  (D) -- (E) node[midway, above] {$\Pi_{56,78}$};;
         \draw (7) -- (D);
        \draw (8) -- (D);
       
        \fill (1)  node[left] {$1$};
        \fill (2) node[below] {$2$};
        \fill (3) node[below] {$3$};
         \fill (4) node[right] {$4$};
          \fill (5) node[right] {$5$};
         \fill (6) node[above] {$6$};
         \fill (7) node[above] {$7$};
         \fill (8) node[left] {$8$};
       \end{tikzpicture}
\caption{The second example of eight-point Witten diagrams with gluon exchanges, where the intermediate bulk-to-bulk gluon propagator $\Pi_{56,78}$ carries nonphysical longitudinal modes.}
\label{fig: 8pt gluon caution}
\end{figure}
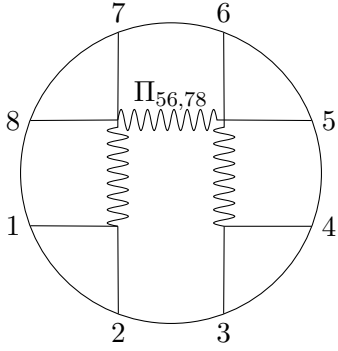
This is an example that is not gauge invariant by itself. The culprit is the bulk-to-bulk gluon propagator $\Pi_{56,78}$ connecting $\phi_7\phi_8$ pair and $\phi_5\phi_6$ pair. To see this, note that the gluon bulk-to-bulk propagators can be modified by adding the gauge-fixing term in the Lagrangian. The gauge-fixing term will bring up spurious longitudinal degrees of freedom. Nevertheless, the amplitudes, as gauge invariant objects, should be independent of the gauge-fixing term. In previous examples, although we only consider a single Witten diagram without summing over to final amplitudes, the gluon bulk-to-bulk propagator always connects two external scalars anti-symmetrically. This anti-symmetry and the on-shell conditions of external legs then ensure the gauge invariance, as the fake longitudinal modes are automatically vanishing by $(L_i-L_j)\cdot (L_i+L_j)=L_i^2-L_j^2\equiv 0$. However, for our example in Fig \ref{fig: 8pt gluon caution}, the longitudinal modes in the propagator $\Pi_{56,78}$ survive because
\be
(L_1-L_2)\cdot (L_1+L_2+L_7+L_8) \neq 0\,.
\ee

\begin{figure}[h]
\centering
 \begin{tikzpicture}
        \draw (0,0) circle (2 cm);
        
        \coordinate (2) at (-0.70,-1.87);
        \coordinate (1) at (-1.87,-0.70);
        \coordinate (A) at (-0.707,-0.707);
        \coordinate (B) at (0.707,-0.707);
        \coordinate (C) at (0,0);
       \coordinate (D) at (-0.707,0.707);
        \coordinate (E) at (0.707,0.707);
         \coordinate (3) at (0.70,-1.87);
        \coordinate (4) at (1.87,-0.707);
         \coordinate (5) at (1.87,0.70);
        \coordinate (6) at (0.70,1.87);
        \coordinate (7) at (-0.70,1.87);
        \coordinate (8) at (-1.87,0.70);
        \draw (1)-- (D);
        \draw (2) -- (D);
          \draw  (3) -- (E);
        \draw (4) -- (E);
         \draw (5) -- (E);
        \draw (6) -- (E);
        \draw[decorate, decoration={coil, aspect=0, segment length=5pt, amplitude=4pt}]  (D) -- (E) node[midway, above] {$\Pi_{56,78}$};;
         \draw (7) -- (D);
        \draw (8) -- (D);
       
        \fill (1)  node[left] {$1$};
        \fill (2) node[below] {$2$};
        \fill (3) node[below] {$3$};
         \fill (4) node[right] {$4$};
          \fill (5) node[right] {$5$};
         \fill (6) node[above] {$6$};
         \fill (7) node[above] {$7$};
         \fill (8) node[left] {$8$};
         \fill (D) circle (2pt);
         \fill (E) circle (2pt);
       \end{tikzpicture}
\caption{The effective amplitude $M^{\rm eff}_{1278g,3456g}$ after performing the Casimir cut $\mathcal{C}_{12}$ and $\mathcal{C}_{34}$.}
\label{fig: 8pt caution eff}
\end{figure}
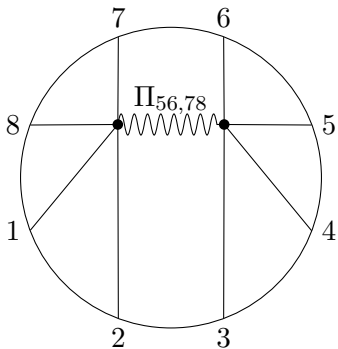

Surprisingly, our method knows that Fig \ref{fig: 8pt gluon caution} is not gauge-invariant by attempting to solve the effective amplitudes $M^{\rm eff}_{1278g,3456g}$ in Fig \ref{fig: 8pt caution eff}.
Naively, we can follow our routine to first make the ansatz
\be
M^{\rm eff}_{1278g,3456g}= \sum_m \fft{f_{1278g}^mP_{ggg}}{\frac{d}{2}+\delta _{12}+\delta _{17}+\delta _{18}+\delta _{27}+\delta _{28}+\delta _{78}-2 \Delta +m-1}\,,\label{eq: ansatz transverse}
\ee
where
\be
& P_{ggg}=2(8\Delta-d)(\delta_{13}-\delta_{14}-\delta_{23}+\delta_{24})C_{ggg}\,,\nn\\
&C_{ggg}=-g_{\rm YM}^4 f^{12a_1}f^{34a_2}(f^{5ba_3}f^{6ba_2}+f^{5ba_2}f^{6ba_3})(f^{7ca_3}f^{8ca_1}+f^{7ca_1}f^{8ca_3}).
\ee
Then we try to solve
\be
\hat{\mathcal{D}}_{1278}^{d-1,1}\circ M^{\rm eff}_{1278g,3456g}=C_{ggg} \big(\hat{L}_{13}-\hat{L}_{14}-\hat{L}_{23}+\hat{L}_{24}\big)\circ 1\,.
\ee
However, we find that there is no solution for $f_{1278g}^m$. This issue is anticipated, as the Witten diagram Fig \ref{fig: 8pt caution eff} cannot be represented by the ansatz \eqref{eq: ansatz transverse}. Instead, \eqref{eq: ansatz transverse} only captures the transverse part of Fig \ref{fig: 8pt caution eff}. Then the question is that how can we obtain the transverse part of the amplitude in Fig \ref{fig: 8pt gluon caution} using our algorithm without knowing its gauge-invariant completion? 

Our trick is to append the possible longitudinal contribution $M_L$, and the correct Casimir equation is
\be
\hat{\mathcal{D}}_{1278}^{d-1,1}\circ M^{\rm eff}_{1278g,3456g} + M_L =C_{ggg} \big(\hat{L}_{13}-\hat{L}_{14}-\hat{L}_{23}+\hat{L}_{24}\big)\circ 1\,.\label{eq: modify eq}
\ee
To proceed, we should make an ansatz for $M_L$. It is worth noting that $M_L$ is not contact, rather it contains all the poles that $M^{\rm eff}_{1278g,3456g}$ have (and probably more). An intuitive way to understand this is to think about its flat-space cousin: the longitudinal mode contributes an additional $1/p^2$. Formally speaking, the longitudinal part of the propagator (after the Casimir cut) gives
\be
\nabla_\mu \nabla^\rho (\Pi_{56,78})_{\rho\nu}\,.
\ee
As the derivatives are moved to the external legs by the integration-by-parts, we are still left with bulk-to-bulk propagators that contribute poles in Mellin variables. Besides, during this procedure, we have a new vertex $V_{1278}$, which can help us make an appropriate ansatz for $M_L$. Ultimately, we find that similar to the flat-space, the longitudinal modes should be proportional to
\be
(L_{1}-L_{2})\cdot(L_1+L_2+L_7+L_8)\times (L_{3}-L_{4})\cdot(L_1+L_2+L_7+L_8)\,,
\ee
motivating an ansatz
\be
M_L= \sum_m \fft{f_{L}^m (\delta_{27}+\delta_{28}-\delta_{17}-\delta_{18})(\delta_{45}+\delta_{46}-\delta_{35}-\delta_{36})}{\frac{d}{2}+\delta _{12}+\delta _{17}+\delta _{18}+\delta _{27}+\delta _{28}+\delta _{78}-2 \Delta +m-1}\,,
\ee
This ansatz allows us to solve \eqref{eq: modify eq} and obtain the results for Fig \ref{fig: 8pt gluon caution} with the only transverse modes
\be
 M_{12g,g78,g56,g34}^{T}=P_{ggg}\sum_{mnp} \ft{f_{ggg}^{mnp}}{(\fft{d}{2}+\delta_{12}-\Delta+m-1)(\fft{d}{2}+\delta_{34}-\Delta+n-1)(\frac{d}{2}+\delta _{12}+\delta _{17}+\delta _{18}+\delta _{27}+\delta _{28}+\delta _{78}-2 \Delta +p-1)}\,,
\ee
where
\be
f_{ggg}^{mnp}=\ft{1}{2(d-2)^2 \Gamma\big(1+4\Delta-\fft{d}{2}\big)} V_{00m}^{1;\Delta-1,\Delta,d-2}V_{00n}^{1;\Delta-1,\Delta,d-2}V_{00pn}^{0;\Delta,\Delta,d-1,d-1}V_{00pm}^{0;\Delta,\Delta,d-1,d-1}S_{m}^{d-2}S_{n}^{d-2}S_p^{d-1}\,.
\ee
\section{Further remarks}
\label{sec: remarks}
\subsection{Relation to the cosmological correlators}

In this paper, we focus on the amplitudes in AdS. Nevertheless, it turns out that the late-time correlators in de Sitter (dS) space, are closely related to AdS amplitudes through analytic continuation \cite{Sleight:2021plv,Sleight:2020obc,DiPietro:2021sjt}. Typically, transitioning from the Euclidean AdS background to dS is rather straightforward; one can simply Wick rotate the AdS radial coordinates to become the dS time, as well as Wick rotate the AdS radius to the Hubble radius. For example, in the Poincaré patch, we have:

\newcommand{\longarrow}[1]{\mkern4.5mu\relbar\mkern-4.5mu\xrightarrow{\hspace*{#1}}}
\be
ds_{\rm AdS} = \frac{R_{\rm AdS}^2}{z^2}(dz^2 + dx^2) \underset{z \rightarrow -it}{\overset{R_{\rm AdS} \rightarrow -i H_{\rm dS}}{\longarrow{2cm}}} ds_{\rm dS} = \frac{H_{\rm dS}^2}{t^2}(-dt^2 + dx^2),.
\ee
As is always the case in quantum field theories, the corresponding analytic continuation of analytic functions, such as scattering amplitudes, is highly nontrivial \cite{eden2002analytic}. We refer readers to \cite{Sleight:2021plv,Sleight:2020obc} for details on analytically continuing the AdS amplitudes to describe the cosmological correlator.

In this section, however, we claim that the Casimir equations for any Witten diagrams remain valid and intact after the analytic continuation (except for switching the signs of those parameters that explicitly depend on the cosmological constant). The reason is that the Casimir equations consist of the boundary conformal generators that are purely defined on either the AdS boundary or the late-time surface in dS, regardless of the analytic properties of the functions these equations solve. Indeed, if we look into the differential representation of four-point functions as in Fig \ref{fig: 4pt Witten} (but with $J=0$), generally we have the Casimir cut
\be
\mathcal{D}^{\Delta,0}_{12} M_{4} = {\rm contact}\,,
\ee
where we should emphasize that $\mathcal{D}_{12}^{\Delta,0}$ is invariant under the shadow $\Delta\rightarrow d-\Delta$. This equation, as Fourier transform to the momentum space, is precisely the cosmological bootstrap equation proposed in \cite{Arkani-Hamed:2018kmz} (specifically see eq (2.41) for $\phi^3$ theory and eq (2.46) for a more general statement in \cite{Arkani-Hamed:2018kmz})! Our exploration in this paper is then expected to provide solid and well-defined cosmological bootstrap equations for higher-point correlators, which may benefit understanding higher spectrum in the slow-roll scenario of the inflation \cite{Arkani-Hamed:2018kmz,Baumann:2022jpr}.

Nonetheless, for the same equation, how do we know whether we are solving the AdS amplitudes or cosmological correlators? This is actually reminiscent of the differential equation method for solving loop Feynman diagrams \cite{Kotikov:1990kg,Remiddi:1997ny,Gehrmann:1999as,Argeri:2007up,Henn:2014qga}, where the Feynman integrals from any analytic contour (e.g., under certain unitary cuts) share the same differential equations. Different contours, as reflected by different analytic properties of the integral results, can be accounted for by the boundary conditions of the differential equations \cite{Anastasiou:2002qz,Anastasiou:2002yz,Henn:2014qga}. (A)dS tree-level correlators follow a similar pattern: we integrate out the bulk coordinates to obtain the amplitudes. The analytic continuation is performed for the bulk coordinates only; therefore, we may find appropriate bulk integral contour in AdS and then deform this contour to achieve the transition from AdS to dS \cite{Sleight:2019hfp,Sleight:2019mgd,Sleight:2021plv}\footnote{Another well-known example is to define the conformal blocks by the geodesic Witten diagram \cite{Hijano:2015zsa,Hijano:2015qja}, where the integral contour is placed to align with the bulk geodesics.}. From the perspective of the Casimir equations, we simply pick up an appropriate boundary condition, as \cite{Arkani-Hamed:2018kmz} did in momentum space. This is a very standard trick when it comes to the analytic continuation of analytic functions, a well-known and most relevant example is the Casimir equation for the conformal block \eqref{eq: eq for conf block}: there are $8$ independent solutions to this Casimir equation with a given $(\Delta,J)$, and they have different asymptotic behaviours; the analytic continuation then picks up specific combinations of these solutions \cite{Caron-Huot:2017vep,Kravchuk:2018htv}. The Casimir equations for (A)dS correlators should follow similarly, for example, $\Delta_+=\Delta$ mixes with its shadow $\Delta_{-}=d-\Delta$ in dS correlators.

However, this argument may raise a puzzle for our algorithm: solutions of Casimir equations are ambiguous, as there are integration constants to be determined by the boundary conditions; however, we find no such constants in the Mellin space. The explanation is that we have already fixed the relevant boundary conditions by making the Mellin ansatz. Our ansatz of Mellin amplitudes is made by the OPE analysis in the unitary CFT, which ensures that we are solving the AdS amplitudes rather than the dS ones. Therefore, a shortcut is that we may be able to make appropriate ansatz for the cosmological correlators for either momentum space \cite{Arkani-Hamed:2018kmz,Baumann:2022jpr}, Mellin space \cite{Penedones:2010ue,Sleight:2020obc} or momentum-Mellin space \cite{Sleight:2019hfp,Sleight:2019mgd} and solve the differential representation unambiguously. For example, it turns out that the correct Mellin ansatz for the four-point cosmological correlators (defined in the Bunch-Davis vacuum) is \cite{Sleight:2020obc}
\be
& M^{\rm dS}_{\nu,J}(s,t)= \fft{1}{2}\sin\big(\fft{\pi}{2}(\Delta_{\Sigma}+2J-d)\big)M^{\rm AdS}_{\Delta_+,J}(s,t)\nn\\
& -\sin\big(\fft{\pi}{2}(\Delta_1+\Delta_2+\Delta_{-}+J-d)\big)
\sin\big(\fft{\pi}{2}(\Delta_3+\Delta_4+\Delta_{-}+J-d)\big) \mathcal{F}_{\nu,J}^{\rm AdS}(s,t)\,,
\ee
where $\Delta_{\pm}=d/2\pm i\nu$ and $\mathcal{F}_{\nu,J}^{\rm AdS}$ is the Mellin transform of the conformal partial waves. In the future, it would be of great interest to understand the cosmological ansatz for any point and then generalize our method to efficiently compute the cosmological correlators to higher points.

\subsection{Flat-space limit}

In the end, we would like to comment on the flat-space limit of the perturbative AdS scattering. This is a kinematic limit where the AdS radius is much larger than any other scales, e.g.,
\be
R_{\rm AdS}\gg b>\ell_s\gg \ell_{pl}\,,
\ee
where $b$ is the impact parameter and $\ell_s$ is the string scale. Intuitively the flat-space limit of AdS observables should give us back their flat-space versions. However, it is generally a hard task to verify and prove such a statement. There are different formalisms for extracting the flat-space observables from AdS amplitudes \cite{Raju:2012zr,Penedones:2010ue,Fitzpatrick:2011hu,Paulos:2016fap,Okuda:2010ym,Maldacena:2015iua,Komatsu:2020sag,Hijano:2019qmi,Hijano:2020szl,Caron-Huot:2021kjy,Cordova:2022pbl,Jain:2022ujj}, see also \cite{ Li:2021snj} for connections between these formalisms, although the underlying analyticity remains subtle \cite{Komatsu:2020sag,vanRees:2022itk} (for extension to supersymmetric amplitudes, see, e.g., \cite{Alday:2017vkk,Rastelli:2016nze,Rastelli:2017udc,Alday:2019nin,Alday:2020dtb,Alday:2021odx}). Surprisingly, the flat-space scattering amplitudes can reversely shed light on the local AdS physics at the AdS Regge limit \cite{Cornalba:2006xk,Antunes:2020pof,Caron-Huot:2021enk}
\be
R_{\rm AdS}\sim b\gg \ell_s\gg \ell_{pl}\,.
\ee
In this limit, flat-space causality constraints can be lifted to AdS with small errors \cite{Caron-Huot:2021enk}.

In this subsection, we emphasize that the differential representation is even more powerful, as it completely uplifts (at least tree-level) the flat-space amplitudes to AdS schematically by replacing the momenta with the conformal generators. For spinning particles, it is also necessary to include the weight-shifting operators \cite{Karateev:2017jgd}, see \cite{Li:2022tby,Lee:2022fgr} for discussions. See also \cite{Bissi:2022wuh} for the differential representation for supersymmetric amplitudes.\footnote{As the differential representation uplifts flat-space amplitudes, it is also expected to provide a formal explanation for the double-copy structures of supersymmetric amplitudes in AdS observed by \cite{Zhou:2021gnu}.} Therefore, the flat-space limit is a corner of this uplifting map, which is now made manifest by the differential representation. The differential representation then provides a more comprehensive way to understand the reason that the flat-space limit should work. We can see this by considering the In\"on\"u-Wigner contraction of the conformal generators \cite{ Giddings:1999jq, Goncalves:2014rfa}
\be
p_\mu = \lim_{R_{\rm AdS}\rightarrow\infty} \fft{1}{R_{\rm AdS}}L_{0\mu}\,,\quad s_{\mu\nu}=L_{\mu\nu}\,,
\ee
where $\mu,\nu$ are indices of the flat-space $\mathbb{R}^{d,1}$; $p_\mu$ and $s_{\mu\nu}$ are the translation and the rotation (with the boost) in $\mathbb{R}^{d,1}$. Therefore, the In\"on\"u-Wigner contraction of the conformal generators gives rise to the Poincare generators in the flat-space \cite{Giddings:1999jq,Goncalves:2014rfa}. Similarly, the In\"on\"u-Wigner of the quadratic conformal Casimir gives rise to the momentum squared in the flat-space
\be
(\sum_{i=1}^n p_i)^2\sim\lim_{R_{\rm AdS}\rightarrow\infty} \fft{1}{R_{\rm AdS}^2}\mathcal{C}_{1\cdots n}\,.
\ee
It is obvious that this provides a group theoretical understanding of the flat-space limit from the differential representation, provided the fact that the $\phi^n$ contact Witten diagram in the flat-space limit is reduced to the momentum conservation delta function \cite{Komatsu:2020sag,Li:2021snj}, because
\be
R_{\rm AdS}^{-2}\, L_i\cdot L_j \rightarrow  p_i\cdot p_j +\mathcal{O}(R_{\rm AdS}^{-2})\,.
\ee
This may provide a systematic way to expand in $1/R_{\rm AdS}$ as a probe to study the Infra-Red (IR) divergence of flat-space S-matrix by viewing $R_{\rm AdS}$ as IR cut-off, e.g., see \cite{Banerjee:2022oll,Duary:2022pyv,Duary:2022afn}. We can also easily understand the flat-space limit in the Mellin space \cite{Penedones:2010ue,Paulos:2016fap}: $L_i\cdot L_j$ acting on the Mellin amplitudes shifts the Mellin variables, at the large scaling limit of the Mellin variables, the action is dominated by multiplying $\delta_{ij}$ to the Mellin amplitudes. We can easily verify all our results in this paper can be reduced to the correct flat-space amplitudes by scaling the Mellin variables $\delta_{ij}\rightarrow\infty$.

\section{Summary}

In this paper, we have proposed a more efficient approach for computing Witten diagrams in AdS using differential representation. Our investigations emphasized that the differential representation is not merely a schematic representation of the amplitudes but should be interpreted as the Casimir equations of Witten diagrams. We used Casimir cuts to visualize the Casimir equations, removing the targeted bulk-to-bulk propagators which provide an iteration to build any tree-level Witten diagram from contact Witten diagrams that are systematically constructed by the boundary Feynman rules. As expected, these boundary Feynman rules directly uplift the flat-space Feynman rules (without higher derivative interactions).

We then explored how we can solve the Casimir equations of Witten diagrams iteratively and efficiently. Our strategy is to study the Witten diagrams in Mellin space, where the differential equations from the Casimir cuts become difference equations in terms of the Mellin variables. In this case, we proposed an algorithm that allows us to bootstrap Witten diagrams in Mellin space from OPE analysis (building the ansatz), analyticity (the cancellation of pole structures), and consistency (matching to the boundary Feynman rules). However, it is worth noting that the term ``bootstrap'' used here differs from the notion of the nonperturbative bootstrap because we are literally doing perturbative calculations and need the Lagrangian to build the diagrams and computing rules. Nevertheless, our method still embodies the spirit of bootstrap by by utilizing symmetry, analyticity, and consistency to obtain results. Using our algorithm, we easily reproduced the four-point Witten diagrams from scalar, gluon, and graviton exchange. Notably, the graviton exchange diagram required efforts to address the contact source terms using standard techniques \cite{Costa:2014kfa}, while our method is more straightforward and efficient.

As for new results, we adapted our algorithm for computing six-point scalar Witten diagrams, where we allowed both the gluon and scalar to be mediated. We found that these amplitudes take a form similar to scalar Mellin Feynman rules \cite{Penedones:2016voo}, but with modified vertex and propagating Gamma functions. Unfortunately, we have not found clean Mellin Feynman rules involving gluons, due to the limited examples. Additionally, we tested our algorithm by solving two examples of eight-point amplitudes involving gluon exchange and demonstrated how to obtain the transverse part of a single non-gauge-invariant Witten diagram.

In the end, we provided non-technical comments on how our methods could be potentially generalized to bootstrap the cosmological correlators in dS, and a simple remark on manifesting the flat-space limit using the differential representation. It is of great interest to follow our routine and compute the cosmological correlators in the future. Besides, it will also be crucial to understand the subtlety of the analyticity in the flat-space limit by using the differential representation.

There are several other important areas for future exploration. One natural extension of our work is to generalize our methods to compute AdS amplitudes at the loop level using the differential representation of loop Witten diagrams \cite{Aharony:2016dwx,Herderschee:2021jbi}. Furthermore, it is crucial to extend our method to compute Witten diagrams involving external spinning particles, such as gluons and gravitons \cite{Li:2022tby}. This investigation will pave the way for understanding the renormalization flow in AdS and its CFT dual \cite{Bertan:2018afl,Aharony:2016dwx,Aprile:2017bgs,Bertan:2018khc,Antunes:2021abs}, and may facilitate the computation of loop observables in AdS, e.g.,\cite{Yuan:2017vgp,Yuan:2018qva,Huang:2021xws,Huang:2023oxf}. In addition, discovering the hidden structure among the gauge theory, such as the BCJ relation and the double copy relation \cite{Bern:2019prr,Bern:2008qj,Bern:2010ue}, would be important for future research. Recent developments in curved space can be found in \cite{Armstrong:2020woi,Farrow:2018yni,Lipstein:2019mpu,Albayrak:2020fyp,Alday:2021odx,Zhou:2021gnu,Bissi:2022wuh,Armstrong:2023phb,Lipstein:2023pih}. Furthermore, our method can likely be applied to supergravity by incorporating the AdS version of SUSY delta functions. This implies a brute force method (other than conformal bootstrap \cite{Gopakumar:2022kof}) to compute the Virasoro-Shapiro amplitude in ${\rm AdS}_5\times {\rm S}_5$ (see \cite{Abl:2020dbx,Aprile:2020mus,Alday:2022xwz,Alday:2022uxp}): expand the Kaluza-Klein reduction and evaluate the diagram term by term.  Finally, our results also provide important ingredients for performing the Mellin bootstrap program \cite{Gopakumar:2016wkt,Gopakumar:2016cpb} for higher-point correlation functions in holographic CFTs, which could shed light on understanding the multi-twist operators \cite{Antunes:2021kmm}.
 
\begin{center}
\textbf{Acknowledgements}
\end{center}
We thank Agnese Bissi, Simon Caron-Huot, Arthur Lipstein, Julio Parra-Martinez, David Simmons-Duffin for useful discussions. We are also grateful to Fei Teng and Xinan Zhou for valuable comments on the draft. YZL is supported by the Simons Foundation through the Simons Collaboration on the Nonperturbative Bootstrap. JM is supported by a Durham-CSC Scholarship.

\begin{appendix}

\section{Boundary Feynman rules}
\label{sec: rules1}
\subsection{Boundary Feynman rules by exchanging gravitons} 
\label{subsec: gravity rules}
We take graviton as example to show how we construct the boundary Feynman rules by addressing the trace of gauge fields. We follow the logic of \cite{Li:2022tby} to present. It turns out that it is instructive to decompose the graviton into traceless part and trace part (with de-Donder gauge) at the level of the Lagrangian.
\be
h_{\mu\nu} = \tilde{h}_{\mu\nu} +\fft{1}{d+1} h g_{\mu\nu}\,,\quad \tilde{h}\equiv 0\,.
\ee
We can then treat $\tilde{h}_{\mu\nu}$ and $h$ separately. In the embedding formalism, they give rise to two different bulk-to-bulk propagators by the wick contractions
\begin{align}
&W_1^\mu W_1^\nu W_2^\rho W_2^\sigma \left\langle \tilde{h}_{\mu\nu}(Y_1)\tilde{h}_{\rho\sigma}(Y_2)\right\rangle = \Pi_{bb}^{d,2}(Y_1,Y_2;W_1,W_2)\,,\nn\\
& \left\langle h(Y_1)h(Y_2)\right\rangle = ({\rm Tr}\,\Pi)_{bb}^{d,2}(Y_1,Y_2)\,.
\end{align}
In the second line, we do not write $W$ dependence to explicitly remind us that there are no polarizations in this case. Typically, when we perform this decomposition, any vertex function coupled to the stress-tensor also effectively makes this decomposition
\be
V_{\mu\nu}h^{\mu\nu} = \tilde{V}_{\mu\nu} \tilde{h}^{\mu\nu} + \fft{1}{d+1} {\rm Tr}\,V\, h\,.
\ee
Without losing the generality, we consider $2$-to-$n$ scalar scattering by exchanging graviton with effective vertices $V^{12}_{\mu\nu}$ and $V^{1\cdots n}_{\mu\nu}$ 
\begin{align}
& W_{2\rightarrow n} = \int D^{d+2}Y_1D^{d+2}Y_2\, \Big(\tilde{V}^{12}_{\mu\nu}\left\langle \tilde{h}^{\mu\nu}(Y_1)\tilde{h}^{\rho\sigma}(Y_2)\right\rangle \tilde{V}^{1\cdots n}_{\rho\sigma}+ \fft{1}{(d+1)^2}{\rm Tr}\,V^{12} \left\langle h(Y_1)h(Y_2)\right\rangle {\rm Tr}\, V^{1\cdots n}\nn\\
&+\fft{1}{d+1}{\rm Tr}\,V^{12}\left\langle h(Y_1)\tilde{h}^{\rho\sigma}(Y_2)\right\rangle \tilde{V}^{1\cdots n}_{\rho\sigma}+\fft{1}{d+1}\tilde{V}^{12}_{\mu\nu}\left\langle \tilde{h}^{\mu\nu}(Y_1)h(Y_2)\right\rangle {\rm Tr}\,V^{1\cdots n} \Big)\,.
\end{align}
Our goal is to show how we can construct the boundary Feynman rules by following the flat-space routine. As we show in subsection \ref{subsec: bulk rules} by using the embedding formalism, we have 
\begin{align}
& \mathcal{D}_{12} W_{2\rightarrow n} = \int D^{d+2}Y_1 D^{d+2}Y_2\times\nn\\
&  \Big(\tilde{V}_{\mu\nu}^{12} (\nabla_{Y_1}^2+2)\left\langle \tilde{h}^{\mu\nu}(Y_1)\tilde{h}^{\rho\sigma}(Y_2)\right\rangle \tilde{V}^{1\cdots n}_{\rho\sigma} + \fft{1}{(d+1)^2} {\rm Tr}\,V^{12}  (\nabla_{Y_1}^2-2d) \left\langle h(Y_1)h(Y_2)\right\rangle {\rm Tr}\, V^{1\cdots n} \nn\\
& +\fft{1}{d+1}{\rm Tr}\,V^{12}(\nabla_{Y_1}^2-2d)\left\langle h(Y_1)\tilde{h}^{\rho\sigma}(Y_2)\right\rangle \tilde{V}^{1\cdots n}_{\rho\sigma}+\fft{1}{d+1}\tilde{V}^{12}_{\mu\nu}(\nabla_{Y_1}^2+2)\left\langle \tilde{h}^{\mu\nu}(Y_1)h(Y_2)\right\rangle {\rm Tr}\,V^{1\cdots n} \Big)\,.\label{eq: trace decomp}
\end{align}
In the next, we have to use the equation of motion in the de-Donder gauge
\be
(\nabla_{Y_1}^2+2) \left\langle h_{\mu\nu}(Y_1)h_{\rho\sigma}(Y_2)\right\rangle-
2g_{\mu\nu} \left\langle h(Y_1)h_{\rho\sigma}(Y_2)\right\rangle=\fft{1}{2}\big(g_{\mu\rho}g_{\nu\sigma}+g_{\mu\sigma}g_{\nu\rho}-\fft{2g_{\mu\nu}g_{\rho\sigma}}{d-1}\big)\delta(Y_1-Y_2)\,.
\ee
It is easy to find that upon the trace decomposition \eqref{eq: trace decomp} we have
\begin{align}
& (\nabla_{Y_1}^2+2) \left\langle \tilde{h}_{\mu\nu}(Y_1)\tilde{h}_{\rho\sigma}(Y_2)\right\rangle=\fft{1}{2}\big(g_{\mu\rho}g_{\nu\sigma}+g_{\mu\sigma}g_{\nu\rho}-\fft{2g_{\mu\nu}g_{\rho\sigma}}{d+1}\big)\delta(Y_1-Y_2)\,,\nn\\
&  (\nabla_{Y_1}^2-2d)\left\langle h(Y_1)h(Y_2)\right\rangle = \fft{2(d+1)}{d-1}\delta(Y_1-Y_2)\,,\nn\\
&  (\nabla_{Y_1}^2-2d)\left\langle h(Y_1)\tilde{h}_{\mu\nu}(Y_2)\right\rangle =  (\nabla_{Y_1}^2+2)\left\langle h_{\mu\nu}(Y_1)h(Y_2)\right\rangle =0\,,
\end{align}
where the first two lines are equivalent to \eqref{eq: traceless eq} and \eqref{eq: trace eq} for massless spin-$2$ field.  Plugging back to \eqref{eq: trace decomp}, it is then straightforward to find
\be
\mathcal{D}_{12}W_{2\rightarrow n} =\int D^{d+2}Y_1 \,V_{\mu\nu}^{12} \times \fft{1}{2}\big(g^{\mu\rho}g^{\nu\sigma}+g^{\mu\sigma}g^{\nu\rho}-\fft{2g^{\mu\nu}g^{\rho\sigma}}{d-1}\big)\times  V_{\rho\sigma}^{1\cdots n}\,,
\ee
namely, the propagator tensor is precisely the same as in flat-space.

\subsection{Summarizing the boundary Feynman rules}

In this section, we summarize the boundary Feynman rules for future reference.  Sticking to the Lorentz gauge, interaction vertex and the Casimir cut of the gluon bulk-to-bulk propagator are given by
\begin{equation}
   \hspace{-5cm}
   \text{Casimir cut of the gluon propagator:}
   \hspace{-2cm}
    \centering
    \begin{minipage}{0.45\linewidth}
        \centering
        \begin{tikzpicture}

        \coordinate (1) at (1,0);
        \coordinate (2) at (-1,0);
        \coordinate (A) at (0,1);
        \coordinate (B) at (0,-1);
            
        \draw[dashed] (B)-- (A);       
        \draw[decorate, decoration={coil, aspect=0, segment length=5pt, amplitude=4pt}]  (1) -- (2) ;

         \fill (2)  node[above] {$\mu$};
         \fill (1)  node[above] {$\nu$};
         
       \end{tikzpicture}
    \end{minipage}
  \hspace{-2cm}
  =g^{\mu\nu}
  \vspace{10pt}
\end{equation}
\begin{equation}
   \hspace{-2cm}
   \text{The conserved current:}
   \hspace{-2cm}
    \centering
    \begin{minipage}{0.45\linewidth}
        \centering
        \begin{tikzpicture}

        \coordinate (1) at (1,0);
        \coordinate (C) at (0,0);
        \coordinate (A) at (-0.7,0.7);
        \coordinate (B) at (-0.7,-0.7);
            
        \draw (B)-- (C);
        \draw (A)-- (C);
        \draw[decorate, decoration={coil, aspect=0, segment length=5pt, amplitude=4pt} ]  (1) -- (C) ;

         \fill (1)  node[above] {$\mu$};
         
       \end{tikzpicture}
    \end{minipage}
  \hspace{-2cm}
  =if^{abc}g_{YM}(\phi_j \nabla_{\mu} \phi_i-\phi_i \nabla_{\mu} \phi_j)
  \vspace{10pt}
\end{equation}

Nest let's consider gravity, for which we take the de-Donder gauge. As we explained in the previous subsection, the corresponding Feynman rules are
\begin{equation}
   \hspace{0.2cm}
   \text{Graviton propagator cut:}
   \hspace{-2cm}
    \centering
    \begin{minipage}{0.45\linewidth}
        \centering
        \begin{tikzpicture}

        \coordinate (1) at (1,0);
        \coordinate (2) at (-1,0);
        \coordinate (A) at (0,1);
        \coordinate (B) at (0,-1);
            
        \draw[dashed] (B)-- (A);       
        \draw[decorate, decoration={snake, aspect=0, segment length=8pt, amplitude=2pt}]  (1) -- (2) ;

         \fill (2)  node[above] {$\mu \nu$};
         \fill (1)  node[above] {$\rho \sigma$};
         
       \end{tikzpicture}
    \end{minipage}
  \hspace{-2cm}
  =\fft{1}{2}\big(g^{\mu\rho}g^{\nu\sigma}+g^{\mu\sigma}g^{\nu\rho}-\fft{2g^{\mu\nu}g^{\rho\sigma}}{d-1}\big)
  \vspace{10pt}
\end{equation}

\begin{equation}
\begin{aligned}
   \hspace{0cm}
   \text{The stress tensor:}
   \hspace{-2cm}
    \centering
    \begin{minipage}{0.45\linewidth}
        \centering
        \begin{tikzpicture}

        \coordinate (1) at (1,0);
        \coordinate (C) at (0,0);
        \coordinate (A) at (-0.7,0.7);
        \coordinate (B) at (-0.7,-0.7);
            
        \draw (B)-- (C);
        \draw (A)-- (C);
        \draw[decorate, decoration={snake, aspect=0, segment length=8pt, amplitude=2pt}]  (1) -- (C) ;

         \fill (1)  node[above] {$\mu \nu$};
         
       \end{tikzpicture}
    \end{minipage}
  \hspace{-2cm}
  =&( \nabla_{\mu}\phi_i \nabla_{\nu} \phi_j+\nabla_{\nu}\phi_i \nabla_{\mu} \phi_j\\
  &-g_{\mu \nu}(\nabla_{\rho}\phi_i \nabla^{\rho} \phi_j+\Delta_{i}(\Delta_{i}-d)\phi_i \phi_j) )
  \vspace{10pt}
  \end{aligned}
  \end{equation}
  
To obtain the final boundary Feynman rules, we need \eqref{eq: from derivative to generator} to rewrite the bulk derivative as boundary differential operators, therefore arriving at the Casimir cut equations for the corresponding Witten diagram. However, we should emphasize that the Casimir cut of the propagator we record here only constructs the correct boundary Feynman rules for the gauge-invariant diagrams, because we literally have fixed the gauge choice. For those Witten diagrams with explicit gauge dependence by themselves, we refer the subsection \ref{subsec: subtle non-gauge} as a prototype for the cautious treatment.

\section{Scalar Mellin Feynman rules}
\label{sec: rules2}
In this section, we provide a brief review of the scalar Mellin Feynman rules, which were summarized in \cite{Penedones:2016voo}. These rules provide a powerful method for calculating Witten diagrams in Mellin space involving scalar fields with zero derivatives.

To apply the Mellin Feynman rules, we first assign a fictitious momentum $p_j$ to every line in the Witten diagram, satisfying the on-shell condition $-p_i^2 = \Delta_i$ for external lines. Momentum is conserved at every vertex of the diagram.

For each internal line, the propagator has a factor given by
\begin{equation}
\frac{S_{m_j}^{\Delta_j}}{p_j^2+\Delta_j+2 m_j},
\end{equation}
where $\Delta_j$ is the dimension of the propagating scalar field, and for internal momentum $p_{ij}=\delta _{ij}$ the corresponding Mellin variables.

The propagator numerator is given by
\begin{equation}
S_m^{\Delta} = \frac{\Gamma(\Delta - \frac{d}{2} + 1 + m)}{2(m!) \Gamma^2(\Delta - \frac{d}{2} + 1)}.
\end{equation}

Finally, the vertex function is given by
\begin{equation}
V_{m_1\ldots m_k}^{\Delta_1\ldots\Delta_k} = \sum_{n_1=0}^{m_1}\cdots\sum_{n_k=0}^{m_k} \Gamma\left(\frac{\sum_j(\Delta_j+2n_j)-d}{2}\right) \prod_{j=1}^k \frac{(-m_j){n_j}}{n_j!(\Delta_j-\frac{d}{2}+1){n_j}}.\label{eq: vertex func}
\end{equation}
Lastly, we sum over all integers $m_j$ from the internal lines, and each sum runs from 0 to $\infty$.

\end{appendix}

\bibliographystyle{JHEP}
\bibliography{refs}

\end{document}